\documentclass[apj]{emulateapj}
\pdfoutput=1
\usepackage{amsmath}
\usepackage{xspace}
\slugcomment{Accepted for publication in ApJ}

\newcommand{\msun}{\ensuremath{\,{\rm M}_{\odot}}\xspace}

\newcommand{\zsun}{\ensuremath{\,{\rm Z}_{\odot}}\xspace}        
\newcommand{\hh}{\ensuremath{\mathrm{H}_2}\xspace}           

\newcommand{\about}{\ensuremath{\sim}}

\newcommand{\RefFig}[1]{\mbox{Figure~\ref{#1}}}                    
\newcommand{\RefEq}[1]{\mbox{Equation~\ref{#1}}}                 
                  
\newcommand{\RefSec}[1]{\mbox{\S~\ref{#1}}}

\begin{document}
\title{The Source Density and Observability of Pair-Instability Supernovae from the First Stars} 
\shorttitle{THE FIRST SUPERNOVAE}
\shortauthors{HUMMEL ET AL.}

\author{ Jacob A. Hummel\altaffilmark{1}, Andreas H. Pawlik\altaffilmark{1},
  Milo\v s Milosavljevi\'c\altaffilmark{1}, Volker
  Bromm\altaffilmark{1,2}} \altaffiltext{1}{Department of Astronomy
  and Texas Cosmology Center, The University of Texas at Austin, TX
  78712} \altaffiltext{2}{Max-Planck-Institut f\"ur Astrophysik,
  Karl-Schwarzschild-Strasse 1, 85740 Garching bei M\"unchen, Germany}

\begin{abstract}
  Theoretical models predict that some of the first stars ended their
  lives as extremely energetic pair-instability supernovae (PISNe).
  With energies approaching $10^{53}$ ergs, these supernovae are
  expected to be within the detection limits of the upcoming {\it
    James Webb Space Telescope} (JWST), allowing observational
  constraints to be placed on the properties of the first stars.  We
  estimate the source density of PISNe using a semi-analytic halo mass
  function based approach, accounting for the effects of feedback from
  star formation on the PISN rate using cosmological simulations. We
  estimate an upper limit of $\sim$0.2 PISNe per JWST field of view at
  any given time.  Feedback can reduce this rate significantly, e.g.,
  lowering it to as little as one PISN per 4000 JWST fields of view
  for the most pessimistic explosion models.  We also find that the
  main obstacle to observing PISNe from the first stars is their
  scarcity, not their faintness; exposures longer than a few times
  $10^4\,$s will do little to increase the number of PISNe
  found. Given this we suggest a mosaic style search strategy for
  detecting PISNe from the first stars. Even rather high redshift
  PISNe are unlikely to be missed by moderate exposures, and a large
  number of pointings will be required to ensure a detection.

\end{abstract}

\section{Introduction}
Understanding the formation of the first stars and galaxies is one of
the central challenges of modern cosmology, as they mark a significant
increase in complexity from the simple initial conditions of the
primordial universe during the `dark ages' (e.g.,
\citealt{BarkanaLoeb2001, Miralda-Escude2003, Brommetal2009,
  Loeb2010}). The basic properties of these so-called Population III
(Pop~III) stars have been reasonably well established, with the
consensus that the first stars formed in dark matter `minihalos' on
the order of $10^5$--$10^6$\msun at high redshifts
\citep{CouchmanRees1986,HaimanThoulLoeb1996,Tegmarketal1997}.
Numerical simulations of the collapse of primordial metal-free gas
into these halos, where molecular hydrogen is the only available
coolant, had suggested that the first stars were predominantly very
massive, with $M_* \gtrsim100$\msun and a top-heavy initial mass
function (IMF) (e.g., \citealt{BrommCoppiLarson1999,
  BrommCoppiLarson2002, AbelBryanNorman2002, BrommLarson2004,
  Yoshidaetal2006, O'SheaNorman2007}).  More recent work has found
that the gas from which the first stars formed underwent significant
fragmentation \citep{StacyGreifBromm2010, Clarketal2011b,
  Greifetal2011, Greifetal2012}, and experienced strong protostellar
feedback \citep{Hosokawaetal2011, StacyGreifBromm2012}.  These results
have revised our picture of Pop~III star formation, with lower
characteristic masses (on the order of 50 rather than $100$\msun) and
a much broader IMF now expected.  The light from these stars ended the
dark ages and fundamentally transformed the universe, beginning both
the reionization (e.g, \citealt{Meiksin2009}) and the chemical
enrichment of the universe (e.g, \citealt{Karlsson2011}).

Given the top-heavy nature of Pop~III star formation and the fact that
low-metallicity stars are unlikely to undergo significant radiatively
driven mass loss \citep{Kudritzki2002}, one interesting possibility is
that some of the first stars died as pair-instability supernovae
(PISNe), a scenario that has significant consequences for the chemical
enrichment history of the universe \citep{HegerWoosley2002,
  TumlinsonVenkatesanShull2004, KarlssonJohnsonBromm2008}.  Basic
one-dimensional models predict that stars with masses in the range
140--260\msun will undergo a pair-production instability and explode
completely \citep{BarkatRakavySack1967, Fraley1968}.  During core
oxygen burning, a combination of high temperatures and relatively low
densities results in the formation of $e^{\pm}$ pairs, removing
pressure support from the core. Following the subsequent contraction,
the ignition of explosive oxygen burning completely disrupts the
progenitor, resulting in a significant contribution to the metal
enrichment of the surrounding medium. More recently,
\citet{ChatzopoulosWheeler2012} and \citet{YoonDierksLanger2012} have
found that stars with initial masses as low as 65\msun can encounter
the pair-production instability if they are rapidly rotating.  In this
scenario, strong rotationally induced mixing causes nearly homogeneous
evolution, such that the star is converted almost entirely to helium
before the next phase in its evolution begins.  These extremely
energetic explosions|approaching $10^{53}$ ergs for the most massive
models|are very luminous, in part due to the large amount of $^{56}$Ni
produced, and are also very temporally extended as a result of the
large mass ejected \citep{FryerWoosleyHeger2001, HegerWoosley2002,
  Hegeretal2003, JoggerstWhalen2011, KasenWoosleyHeger2011}.

A second possiblility for reaching such extreme explosion energies in
slightly lower mass stars is the hypernova scenario for rapidly
rotating stars that undergo core collapse \citep{UmedaNomoto2003,
  TominagaUmedaNomoto2007}.  During the collapse, accretion onto a
central black hole powers a jet which induces a highly energetic
explosion. While recent work has decreased the expected mass of the
first stars, they have also been found to rotate more rapidly than
previously thought \citep{StacyBrommLoeb2011}, increasing the
plausibility of this scenario. Evidence supporting this hypothesis has
recently been presented by \citet{Chiappinietal2011}.

While not an example of a Pop~III star, the recent discovery of the
extremely luminous supernova (SN) 2007bi, identified as a possible
PISN, in a metal-poor dwarf galaxy at a redshift of $z\simeq0.1$
\citep{Gal-Yametal2009} suggests that PISNe may be possible in the
local universe under rare circumstances. Further supporting this
picture, \citet{WoosleyBlinnikovHeger2007} have shown that SN 2006gy
\citep{Smithetal2007} is well modeled by a pulsational
pair-instability model. For stars with initial masses in the range
$\sim$100--140\msun, the star encounters the $e^{\pm}$ production
instability, but the resulting explosive ignition of oxygen is
insufficient to unbind the star.  Instead it ejects a shell of
material before settling back into a stable configuration.  The star
encounters this instability several times until the mass of the helium
core drops below $\sim$40\msun, after which the star can proceed to
silicon burning and eventually undergo core-collapse.

With the upcoming launch of the {\it James Webb Space Telescope}
(JWST) we will be able to probe the epoch of first light in
unprecedented detail.  While the first stars themselves are unlikely
to be visible (e.g, \citealt{BrommKudritzkiLoeb2001,
  PawlikMilosavljevicBromm2011}), some of the SNe that end their lives
should be within the detection limits of the JWST (e.g.,
\citealt{MackeyBrommHernquist2003, Scannapiecoetal2005,
  Gardneretal2006}). While the basic properties of PISNe and the
effect they have on their environment has been well studied
\citep{MoriFerraraMadau2002, BrommYoshidaHernquist2003,
  FurlanettoLoeb2003, KitayamaYoshida2005, Whalenetal2008,
  WiseAbel2008, Greifetal2010}, the source density of these events has
yet to be well constrained.

The first attempt at estimating the number and observability of SNe at
high redshift was made by \citet{Miralda-EscudeRees1997}, who
calculated the all-sky SN rate based on estimates of the total metals
produced by a typical SN and the observed metallicity of the
intergalactic medium (IGM) at high redshifts.  This yields $\sim$1 SN
yr$^{-1}$ arcmin$^{-2}$ at $z \sim 5$. Other early work attempted to
model the SN rate based on the empirically determined star formation
rate out to high redshifts ($z \sim 5$; e.g.,
\citealt{MadauDellaVallePanagia1998, DahlenFransson1999}).  It should
be noted that these attempts focused on Type II SNe, not PISNe, which
are the focus of the present work. \citet{MackeyBrommHernquist2003}
estimated the PISN rate based on their calculations of the Pop~III
star formation rate, predicting $\sim$$2\times10^6$ PISNe yr$^{-1}$
over the whole sky above $z=15$.  \citet{WeinmannLilly2005} performed
a similar analysis with more conservative estimates for the star
formation rate, finding a PISN rate of $\sim$4 yr$^{-1}$ deg$^{-2}$
above $z=15$ and $\sim$0.2 yr$^{-1}$ deg$^{-2}$ above $z=25$, as well
as concluding that PISNe should be observable out to $z=50$ with the
JWST.

Subsequent work by \citet{WiseAbel2005} determined the PISN rate based
on the collapse of gas into dark matter minihalos. Accounting for
radiative feedback, they concluded that $\sim$0.34 PISNe yr$^{-1}$
deg$^{-2}$ were to be expected above $z=10$, as well as briefly
considering the detectability of PISNe at high redshifts based on
models from \citet{HegerWoosley2002}.  \citet{Scannapiecoetal2005}
presented a more thorough analysis of the visibility of PISNe based
on a suite of numerical simulations spanning the range of theoretical
PISN models using the implicit hydrodynamics code
KEPLER. \citet{MesingerJohnsonHaiman2006} presented a similar but more
general halo mass function based analysis, considering the rates and
detectability of all SNe; they briefly consider primordial PISNe, but
focus on core-collapse SNe and their detectability.  More recently,
\citet{TrentiStiavelliShull2009} explored the observable PISN rate
within the context of the metal-free gas supply during the epoch of
reionization.

Our work improves upon these investigations by incorporating updated
PISN models from \citet{KasenWoosleyHeger2011} and determining their
observability using the published specifications of the Near Infrared
Camera (NIRcam) on the
JWST\footnote{http://www.stsci.edu/jwst/instruments/nircam}.
In the final stages of the work on this paper we have become aware of
the study by \citet{PanKasenLoeb2012}, who have performed a similar
analysis. This paper addresses the question of PISN observability in a
nicely complementary way by employing a different normalization
strategy. Different from our assumption that viable Pop~III
progenitors can only form in unenriched minihalos at $z\gtrsim6$,
\citet{PanKasenLoeb2012} derive the star formation rates required to
produce sufficient photons for reionization.  They then infer the PISN
rate corresponding to different choices for the IMF.  Their analysis
is thus able to probe the charachter of star formation in the dwarf
galaxies that are the main drivers of reionization, whereas we focus
on the minihalos where Pop~III stars first begin to form.

This paper is organized as follows.  In Section 2 we describe our
semi-analytic model for the PISN rate.  We consider the ability of the
JWST to detect PISNe at high redshift in Section 3, and our
conclusions are gathered in Section 4.  Throughout this paper we adopt
a $\Lambda$CDM model of hierarchical structure formation, using
cosmological parameters consistent with the WMAP 5-year results
\citep{Komatsu2009}: $\Omega_{\rm m} = 0.258$; $\Omega_{\Lambda}=0.742$;
$\Omega_b=0.0441$; $h=0.719$; $n_s=0.96$; $\sigma_8=0.796$.

\section{The PISN Rate}
PISNe are produced only by very massive stars
\citep{BrommKudritzkiLoeb2001, Schaerer2002, Hegeretal2003}, which are
now expected to be rare even for Pop~III star formation.  After the
first massive star forms, the resulting heating from photoionization
quickly suppresses the density of the remaining gas in the minihalo,
effectively halting star formation \citep{Kitayamaetal2004,
  WhalenAbelNorman2004, AlvarezBrommShapiro2006}. The energy released
by the first PISN disperses the gas in the halo and contaminates it
with metals \citep{BrommYoshidaHernquist2003, Greifetal2007,
  Whalenetal2008, WiseAbel2008, Greifetal2010}.  Subsequent episodes
of star formation are thus delayed until the gas is able to recondense
into more massive cosmological halos \citep{YoshidaBrommHernquist2004,
  Yoshidaetal2007, JohnsonGreifBromm2007, AlvarezWiseAbel2009}.  While
star formation will resume at this point, the gas in these systems is
expected to be enriched beyond the critical metallicity for the
transition to Population II (Pop~II) star formation \citep{WiseAbel2007, WiseAbel2008,
  Greifetal2007, Greifetal2008, Greifetal2010}. As a result, the stars
that form will no longer be massive enough to reliably produce PISNe.
These explosions are thus only expected to occur in minihalos
containing pristine gas that has just crossed the density threshold
for star formation via H$_2$ cooling, and only one PISN occurs per halo.

It is possible that the first stars formed in binaries or small
multiples; however, the number of massive stars formed per minihalo is
still of order unity \citep{StacyGreifBromm2010, Clarketal2011b,
  Greifetal2011}.  Given this, and assuming that the time required for the
progenitor star to form, live and die is negligible
\citep{Hegeretal2003}, we can use the formation rate of minihalos to
place a robust upper limit on the PISN rate.

The introduction of cosmic feedback has the potential to significantly
alter this picture.  Chemical enrichment induces a transition to lower
mass Pop~II star fomation, and thus always has a negative effect on
the PISN rate.  Radiative feedback---especially \hh-dissociating
Lyman-Werner (LW) feedback---has a more complicated effect.  The
destruction of molecular hydrogen by LW photons removes the ability of
pristine gas to cool effectively. This suppresses star formation, and
hence negatively affects the PISN rate.  However this merely delays
star formation; as the halo mass continues to increase, the gas
eventually becomes self-shielding and proceeds with cooling and
collapse.  The delay effectively increases the amount of gas available
when star formation begins. This increases the possibility of multiple
massive stars forming simultaneously, positively affecting the PISN
rate.  To account for these possibilities, we consider three
scenarios.  First we estimate the PISN rate assuming that the first
stars form unimpeded by cosmic feedback. Then we consider a
conservative scenario in which both chemical and LW feedback affect
the PISN rate, but only one star forms per halo.  Finally, we allow
for enhanced massive star formation in the LW-affected halos, and calculate
the observable rate for each scenario.

\subsection{No-Feedback Limit}
In order to determine an upper limit to the PISN rate, we assume
exactly one PISN per minihalo, forming as soon as the virial
temperature of the minihalo exceeds the minimum value $T_{\rm crit}$
required for gas to cool and collapse to high densities. We set this
to 2200 K based on the results of simulations (see
\RefSec{lymanWerner} for details).  The corresponding critical mass
for collapse is given by
\begin{equation}
  \label{mcriteq}
    M_{\rm crit} = 10^6\msun
    \left( \frac{T_{\rm crit}}{10^3 \,{\rm K}}\right)^{3/2} 
   \left( \frac{1+z}{10}\right)^{-3/2},
\end{equation}
where we assume a mean molecular weight of $\mu=1.22$, appropriate for
the almost completely neutral IGM at high redshifts
\citep{BarkanaLoeb2001}.

We use the analytic Press-Schechter (PS) formalism for structure
formation \citep{PressSchechter1974} to estimate the number density
$n_{\textsc{ps}}$ of minihalos of mass $M$ at redshift $z$, given by
\begin{equation}
n_{\textsc{ps}} (M,z) = \sqrt{\frac{2}{\pi}}
\frac{\rho_{\rm m}}{M}\left|\frac{d\ln \,\sigma_{0}(M)}{d\ln \,M}
\right|\nu_{c,z} \,e^{-\nu_{c,z}^2 /2},
\end{equation}
where $\rho_{\rm m}$ is the background matter density, $\sigma_{z}(M)$
is the standard deviation of overdensities $\delta$ of mass $M$ at
redshift $z$, and $\nu_{c,z} = \delta_c / \sigma_{z}(M)$, where
$\delta_c$ is the critical overdensity for collapse; the value used
here is $\delta_c =1.686$.

Converting from redshift to cosmic time $t(z)$, where
\begin{equation} 
t(z) = \frac{1}{H_0}
\int_z^{\infty}\frac{dz'}{(1+z')\sqrt{\Omega_{\rm m} (1+z')^3 + \Omega_{\Lambda}}} ,
\end{equation} 
a first order estimate for the PISN formation rate as a function
of redshift is given by $\dot{n}_{\textsc{pisn}}\equiv
dn_{\textsc{ps}}/dt$. However, this estimate is only valid while the
rate of destruction of minihalos via mergers remains small compared to
the formation rate. The Press-Schechter formalism only gives the total
number of halos of a given mass at a given redshift. As a result, once
mergers become important the rate of change of the total number of
minihalos no longer traces the formation rate. To correct for this, we
use the expression for the formation rate derived by
\citet{Sasaki1994}\footnote{See \citet{Mitraetal2011}
  for a discussion of the validity of this expression.}:
\begin{equation}
\dot{n}_{+}(z) = \frac{\dot{D}}{D} n_{\textsc{ps}}(M_{\rm crit},z)
\frac{\delta_c^2}{\sigma^2_0(M) D^2},
\end{equation}
where $D(z)$ is the growth factor.  The PISN rate
in this upper limit of no feedback, shown in \RefFig{fbrate} (blue
line), is then simply given by the halo formation rate:
\begin{equation}
\dot{n}_{\textsc{pisn}}(z) = \dot{n}_{+}(z).
\end{equation}

\subsection{Feedback}
The preceding analysis has only nominally incorporated the baryonic
physics involved through the critical mass for H$_2$ cooling. Gas will
not successfully cool and collapse in all minihalos that reach the
critical mass (e.g., \citealt{Yoshidaetal2003}). The various feedback
mechanisms responsible for this include photoheating from stars in
nearby halos and the buildup of a background of H$_2$ dissociating LW
photons. Chemical feedback will enrich the gas with metals, improving
its ability to cool.  However, gas that is enriched forms lower-mass
Pop~II stars, effectively reducing the PISN rate. These
feedback mechanisms can be represented with distinct efficiency
factors $\eta(z)$, such that the true PISN rate will be given by
\begin{equation}
  \dot{n}_{\textsc{pisn}}(z) = 
  \eta_{\rm chem}(z) \; \eta_{\rm rad}(z) \; \dot{n}_{+}(z).
  \label{pisnrate1}
\end{equation}
We must include these effects in order to derive a realistic estimate
for the PISN rate, which we henceforth refer to as the conservative
feedback case.

\subsubsection{Lyman-Werner Feedback}
\label{lymanWerner}
LW feedback is of particular importance in any discussion of feedback
on the first stars as it dissociates the H$_2$ molecules primarily
responsible for cooling primordial gas.  This significantly reduces
the ability of the gas to cool and works to suppress further star
formation \citep{HaimanReesLoeb1997, OmukaiNishi1999,
  CiardiFerraraAbel2000, HaimanAbelRees2000, GloverBrand2001,
  Kitayamaetal2001, MachacekBryanAbel2001, RicottiGnedinShull2001,
  RicottiGnedinShull2002a, RicottiGnedinShull2002b, Yoshidaetal2003,
  OmukaiYoshii2003, MesingerBryanHaiman2006}.  As the first stars form
in $\gtrsim 3\sigma$ peaks in the Gaussian distribution of density
fluctuations \citep{BarkanaLoeb2001}, we focus here on a similarly
overdense environment in order to provide a conservative estimate of
the effects of LW feedback on the PISN rate.  While representative of
the LW background in Pop~III star formation sites, this is likely not
representative of the `average' LW background in the universe (e.g.,
\citealt{MachacekBryanAbel2001, MesingerBryanHaiman2006}).

\begin{figure}
 \begin{center}
   \includegraphics[width=\columnwidth]{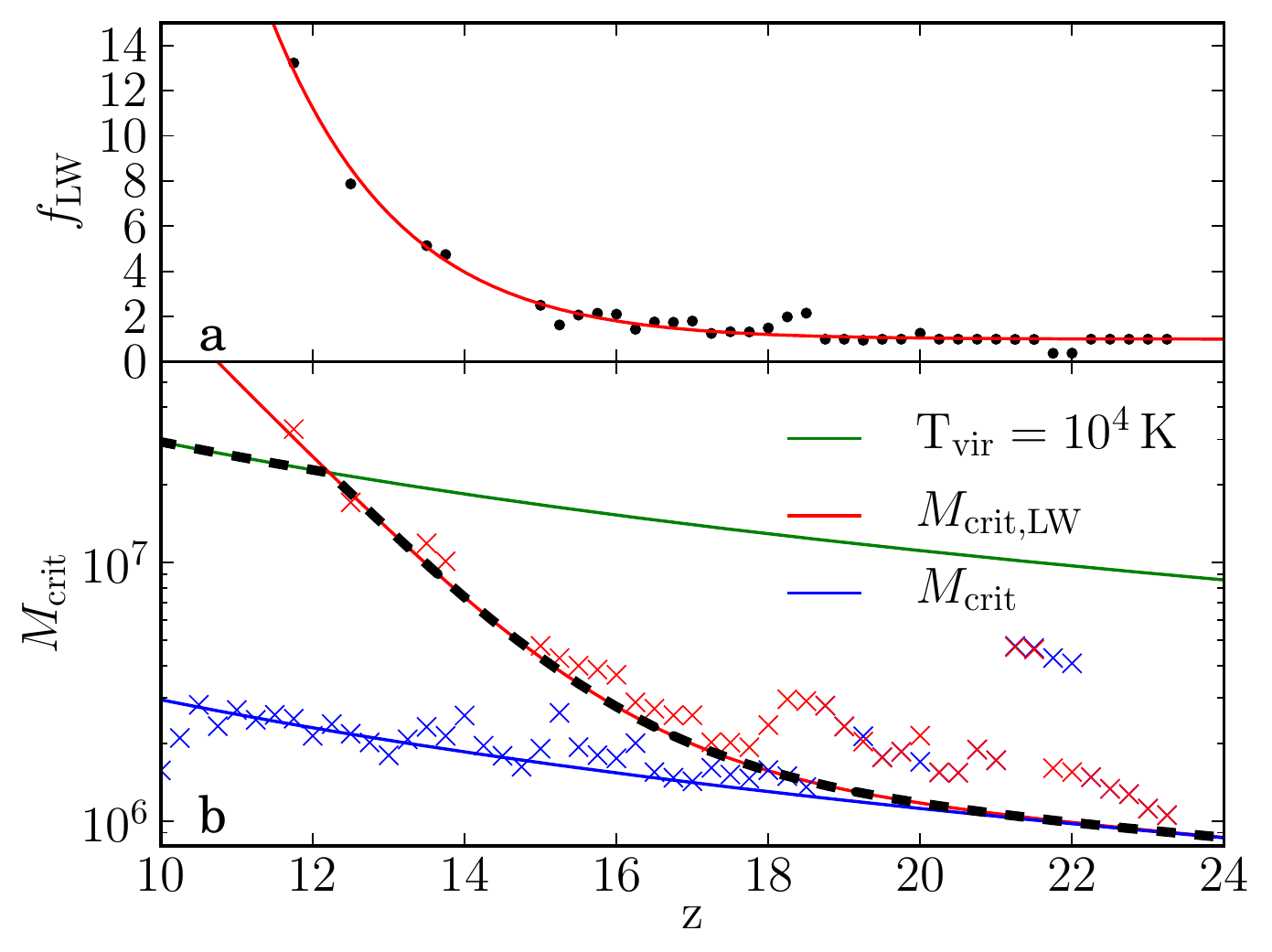}
   \caption{\footnotesize a) the factor $f_{\textsc{lw}}$ by which
     LW feedback increases the critical mass for star formation;
     points show the results of the simulation, the red line our fit.
     b) The critical mass for star formation in both simulations.
     Blue crosses mark the critical mass for star formation in the
     absence of LW feedback, red crosses in its presence.  The blue
     line shows the best fit critical mass from \RefEq{mcriteq}, given
     by a temperature of 2200 K; the red line the critical mass from
     \RefEq{mcritlweq} for LW feedback using the fit for
     $f_{\textsc{lw}}$ given in \RefEq{flweq}.  The green line marks
     the halo mass corresponding to a virial temperature of $10^4\,$K,
     and the black dashed line represents the critical mass employed,
     accounting for atomic cooling in halos with virial temperatures
     above $10^4\,$K.}
   \label{lwFeedback}
 \end{center}
\end{figure}

We will describe our radiative feedback simulations in detail
elsewhere, and present only a brief summary here.  We employ a set of
two cosmological simulations using a modified version of the
$N$-Body/TreePM SPH code GADGET \citep{Springel2005,Springeletal2001}
to gauge the effects of LW feedback on the PISN rate. These
simulations, carried out in a box of size $3.125 \,h^{-1}$ comoving
Mpc and starting from a redshift of $z=127$, were performed to
investigate the formation of the first dwarf galaxies in $10^9$\msun
halos at $z=10$. In order to obtain high resolution in the halo
containing the first galaxy while retaining information about
structure on large scales, a zoomed simulation technique was used,
with the highest resolution dark matter (gas) particles having a mass
of 2350 (484)\msun. This allows halos with masses $\gtrsim 2
\times 10^5$\msun, to be resolved with $\gtrsim 100$ dark
matter particles.

The first of the simulations we employ is similar to simulation {\it
  Z4} presented in \citet{PawlikMilosavljevicBromm2011}, and follows
the non-equilibrium chemistry and cooling of the primordial atomic and
molecular gas, including star formation but not the associated
feedback.  It thus provides a useful reference against which the
effects of LW feedback can be discussed. Here, gas particles were
turned stochastically into star particles at densities $n_{\rm H} >
500 \,{\rm cm}^{-3}$ on a dynamical time scale. Star particles were
considered simple stellar populations using the zero-metallicity
top-heavy IMF models from \citet{Schaerer2003}.  Henceforth this
simulation is referred to as Simulation A.

The second simulation employed here, which we refer to as Simulation
B, is identical to Simulation A except for the inclusion of LW
feedback.  Calculation of the feedback was carried out by considering
the contribution from both star particles and a uniform LW background.
The combined LW background is normalized to approximate the LW
background evolution shown in \citet{GreifBromm2006} in the optically
thin limit, but with the application of a self-shielding correction
\citep{Wolcott-GreenHaimanBryan2011}.

The efficiency of LW feedback, $\eta_{\textsc{lw}}$, can be expressed
as the ratio of the formation rate of minihalos at the critical mass
with LW feedback to that without:
\begin{equation}
  \eta_{\textsc{lw}}(z) = \frac{ \dot{n}_{+}(M_{\rm crit, \textsc{lw}},
    z) }{ \dot{n}_{+}(M_{\rm crit}, z)}.
\end{equation}
This requires determining the factor $f_{\textsc{lw}}(z)$ by which LW
feedback increases the critical mass for star formation.  We do this
by determining the critical mass required for stars to form in both
simulations, as traced by the lowest mass halo that is actively
forming stars for the first time.  The resulting critical mass in each
case is shown in \RefFig{lwFeedback}b; blue points denote the critical
mass in Simulation A, red points in Simulation B.  Without any
negative feedback, we expect stars to form when the halo reaches the
critical mass given in \RefEq{mcriteq}.  We find the that the resulting critical
mass in Simulation A is best fit by a critical virial temperature of
2200 K, as shown in \RefFig{lwFeedback}b.  Note that this includes the
effects of dynamical heating; gas in isolated halos would collapse and
form stars at lower masses.

\begin{figure}
 \begin{center}
   \includegraphics[width=\columnwidth]{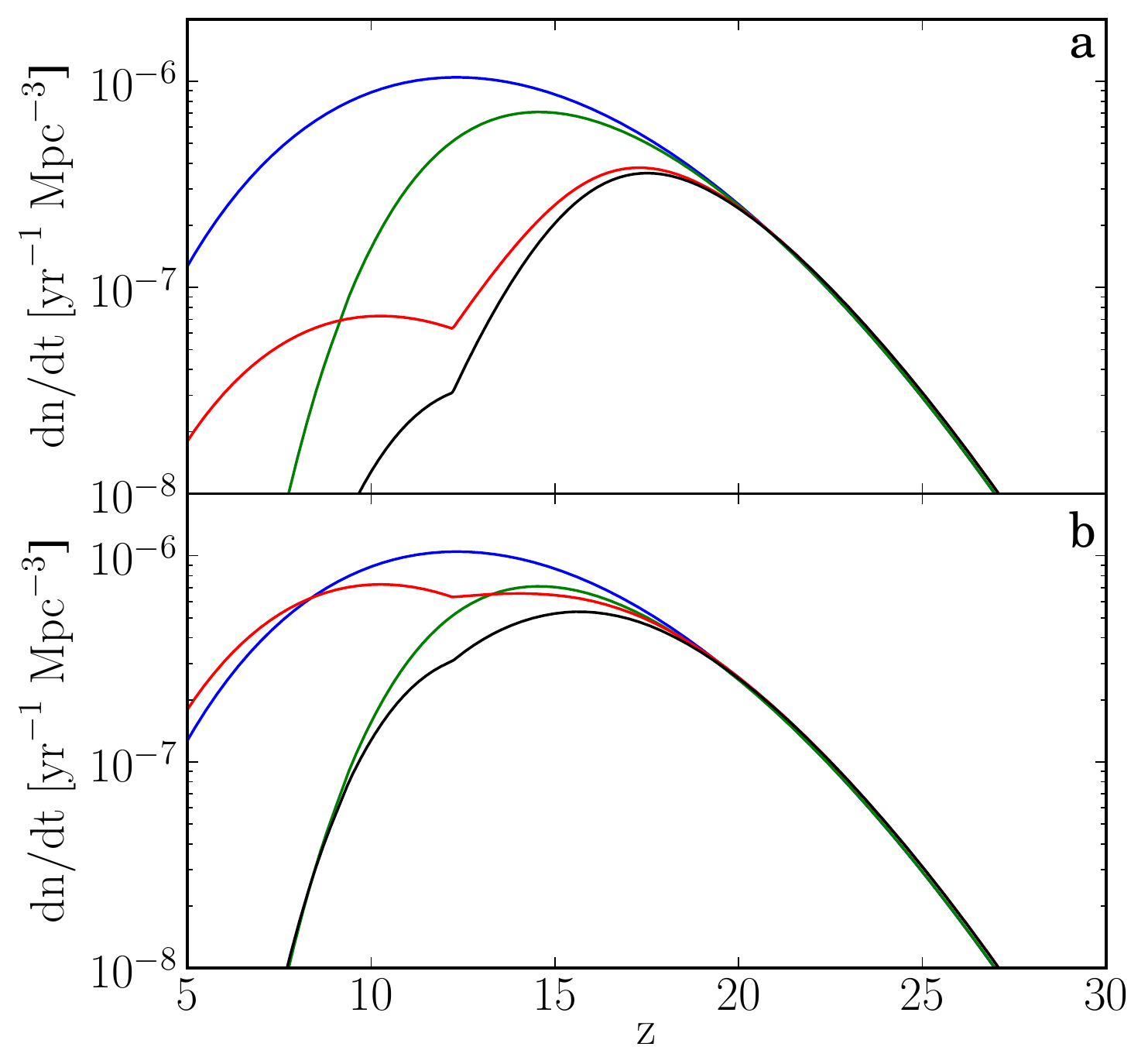}
   \caption{a) $\dot{n}_{\textsc{pisn}}$ in the upper limit of no
     feedback (blue), with chemical feedback (green), LW feedback
     (red) and the resulting PISN rate for the conservative (chemical
     plus LW) feedback case (black).  b) Same as (a), but for enhanced
     massive star formation.}
   \label{fbrate}
 \end{center}
\end{figure}

Determining the critical mass for star formation in the LW feedback
simulation in the same manner as above, we can then determine
$f_{\textsc{lw}}$, shown in \RefFig{lwFeedback}a.  We find that
$f_{\textsc{lw}}$ is well fit by a functional form of
\begin{equation}
  \label{flweq}
  \begin{split}
 f_{\textsc{lw}}(z) = -6.23\times10^{-5} \;
 &\textrm{erf}[0.094 (z - 0.204)] \\
 &+ 6. 23\times10^{5},
 \end{split}
\end{equation}
for $z\geq10$, where ${\rm erf}(z)$ is the error function.
$M_{\rm crit, \textsc{lw}}$ is then given by
\begin{equation}
  \label{mcritlweq}
M_{\rm crit, \textsc{lw}} = f_{\textsc{lw}} \,M_{\rm crit},
\end{equation}
shown in \RefFig{lwFeedback}b.  When the virial temperature of the
halo reaches $10^4\,$K, cooling via atomic hydrogen becomes efficient
and molecular hydrogen becomes unimportant.  Once this threshold is
reached, LW feedback cannot further suppress star formation, and
$M_{\rm crit, \textsc{lw}}$ holds steady at a constant virial
temperature of $10^4\,$K.  This is reflected as an increase in the
formation rate of halos affected only by LW feedback, as the formation
rate of atomic cooling halos is increasing prior to $z=10$.  This
transition can be clearly seen in \RefFig{fbrate}, marked by a sudden
jump in the halo formation rate.  As ionizing photons cannot easily
escape the immediate vicinity of the star that produced them, their
impact on neighboring halos is small compared to that of LW photons at
the highest redshifts.  We thus set $\eta_{\rm rad} \simeq
\eta_{\textsc{lw}}$ for simplicity.  This approximation becomes
increasingly unphysical close to the epoch of reionization at
$z\sim10$.

\subsubsection{Chemical Feedback}
The process of chemical enrichment is another crucial factor for
determining the PISN rate.  Gas that has been enriched beyond a
critical metallicity of $Z_{\rm crit} \sim 10^{-4}$\zsun will
no longer form Pop~III stars \citep{BrommKudritzkiLoeb2001,
  Schneideretal2002, BrommLoeb2003}, and hence no PISNe.  Chemical
feedback can thus be represented as the fraction of halos forming from
pristine gas at a given redshift.  Realistic three-dimensional
simulations of this process starting from cosmological initial
conditions have become possible in the past decade, showing that
enrichment by Pop~III SNe, if they are highly energetic, proceeds very
inhomogeneously, enriching the IGM before penetrating into denser
regions \citep{Scannapiecoetal2005, Greifetal2007,
  TornatoreFerraraSchneider2007,WiseAbel2008, Maioetal2010}.

In modeling $\eta_{\rm chem}$, we use the results of
\citet{FurlanettoLoeb2005}.  Their semi-analytic treatment of SN winds
utilizes the \citet{Sedov1959} solution for an explosion expanding
into a uniform medium and yields a probability function $P_{\rm
  pristine}(z)$ that the gas in a newly formed halo is pristine.  This
is plotted in Figure 2 of their paper for various strengths of
chemical feedback.  We identify this quantity as the fraction of newly
collapsed halos that have been polluted with metals, $\eta_{\rm
  chem}$.  Given the recent detection of pristine gas at $z = 3$ by
\citet{FumagalliOMearaProchaska2011}, we choose the weakest feedback
scenario presented by \citet{FurlanettoLoeb2005} among the scenarios
that incorporate a clustering of sources. The resulting PISN rate is
given by the green line in \RefFig{fbrate}.

\subsection{Enhanced Massive Star Formation}
Gas cooling and subsequent star formation in halos affected by LW
feedback can be delayed until nearly an order of magnitude more gas is
available for star formation (\RefFig{lwFeedback}). This increases the
likelihood that multiple massive stars form per halo, offseting the
negative effects of LW radiation considered above.  We quantify this
by positing that the number of PISNe produced per halo at redshift $z$
is given by the ratio of the critical mass in the presence of LW
feedback $M_{\rm crit, \textsc{lw}}$ to the critical mass in the
no-feedback case $M_{\rm crit}$.  For example, at $z=17$, $M_{\rm
  crit, \textsc{lw}} / M_{\rm crit} \approx 1.4$, so for every 10
pristine halos that form, 14 PISNe are produced.  In this case the
PISN rate is modified such that
\begin{equation}
\dot{n}_{\textsc{pisn}}(z) = \frac{M_{\rm crit, \textsc{lw}}(z)}{M_{\rm crit}(z)} \;
\eta_{\rm chem}(z) \; \eta_{\rm rad}(z) \; \dot{n}_{+}(z).
\label{pisnrate2}
\end{equation}
The resulting enhanced PISN rate can be seen in \RefFig{fbrate}b.  In
contrast to the conservative feedback case, the net effect of LW
feedback is much less significant here, with chemical feedback
controlling the final PISN rate.

\subsection{The Observable Rate}
\begin{figure}
 \begin{center}
   \includegraphics[width=\columnwidth]{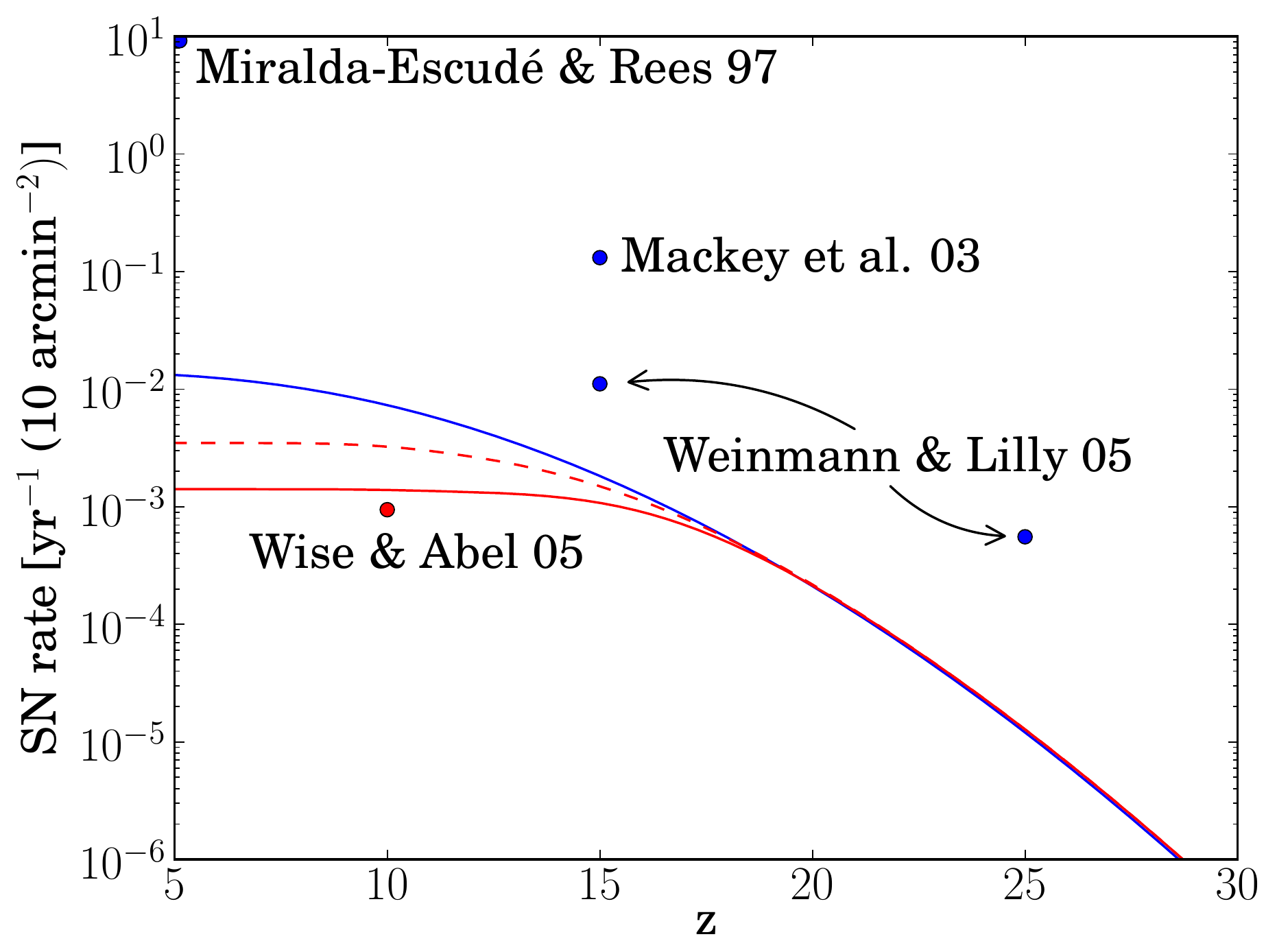}
   \caption{\footnotesize The observable PISN rates in number per year
     per JWST field of view above a given redshift in the upper limit
     of no feedback (blue line), in the conservative feedback case
     (solid red line), and the enhanced star formation case (dashed
     red line).  The rates calculated by
     \citet{Miralda-EscudeRees1997}, \citet{MackeyBrommHernquist2003},
     \citet{WeinmannLilly2005} and \citet{WiseAbel2005} are also shown
     for reference. Red points account for feedback; blue points do
     not.}
   \label{obsrate}
 \end{center}
\end{figure}
The observed PISN rate per unit time per unit redshift per unit solid
angle is given by
\begin{equation}
  \begin{split}
    \frac{dN}{dt_{\rm obs}\,dz\,d\Omega} &= \frac{dN}{dt_{\rm
        obs}\,dV} \, \frac{dV}{dz d\Omega}
    \\ &=\frac{1}{(1+z)}\frac{dN}{dt_{\rm em}\,dV} \;r^2
    \frac{dr}{dz}.
 \end{split}
\end{equation}
Cosmological time dilation between $t_{\rm obs}$ and $t_{\rm em}$ is
accounted for by the $(1+z)$ in the denominator; $dV$ is the
comoving volume element and $r(z)$ is the comoving distance to
redshift $z$ given by
\begin{equation}
r(z) = \frac{c}{H_0}\int_0^z \frac{dz'}{\sqrt{\Omega_{\rm m} (1+z')^3 +
    \Omega_{\Lambda}}},
\end{equation}
where $c/H_0$ is the Hubble distance.  With the assumptions outlined
above, we estimate the PISN rate in events per year per comoving
Mpc$^3$ in the source rest frame:
\begin{equation}
\frac{dN}{dt_{\rm em}\,dV} = \dot{n}_{\textsc{pisn}}(z).
\end{equation}
These results---shown in \RefFig{obsrate}---are in reasonable
agreement with previous work; our no-feedback limit of one PISN per
minihalo is somewhat more conservative than that employed by
\citet{WeinmannLilly2005}, but in general agreement.  Likewise, our
conservative feedback rate is in good agreement with the rate found by
\citet{WiseAbel2005}, which also accounted for feedback.  The
discrepancy with the remaining rates can be attributed to the fact
that \citet{MackeyBrommHernquist2003} employed an optimistic star
formation efficiency of $\eta_*=0.10$, while
\citet{Miralda-EscudeRees1997} performed a rough estimate of the
all-sky supernova rate based on the observed metallicity of the
IGM. This estimate of course includes Population~I and II supernovae in
addition to PISNe, and is included only for reference.

\section{JWST Observability}
\label{JWSTobs}
While PISN explosions are predicted to be extremely energetic, the
highest redshift events will still be unobservable, and those at lower
redshifts will be above the detection limits of the JWST for only a
fraction of their lifetimes. To determine the observability of these
explosions we must consider both how bright they will be at a given
redshift and how long they will remain visible.  To span the
uncertainties arising from variation in the progenitors of PISNe we
consider a set of four models from \citet{KasenWoosleyHeger2011},
namely their R250, B200, R175 and He100 models.  R250 and R175 are red
supergiants of 250\msun and 175\msun, respectively, spanning the mass
range of succesful explosions.  We also consider a more compact
200\msun blue supergiant (B200) and a 100\msun bare helium core
(He100).  All models considered here die as PISNe, with explosion
energies ranging from $7\times10^{52}$ ergs (R250) to $2\times10^{52}$
ergs (R175).  The late-time luminosity is powered by the decay of
$^{56}$Ni produced during the explosion.  40\msun of $^{56}$Ni are
produced in the R250 model, while only 5, 2 and 0.7\msun are produced
in He100, B200 and R175, respectively.

Bare helium cores, such as the He100 model, are of particular interest
in light of recent work finding that rapidly rotating stars can
encounter the pair-production instability in progenitors with masses
as low as 65\msun \citep{ChatzopoulosWheeler2012,
  YoonDierksLanger2012}. Combined with recent work finding that
Pop~III stars are both less massive and more rapidly rotating than
previously thought \citep{StacyGreifBromm2010, StacyBrommLoeb2011,
  StacyGreifBromm2012, Clarketal2011b, Greifetal2011, Greifetal2012},
the explosion of a rapidly rotating helium core formed by homogeneous
evolution represents an intriguing possibility.

We first describe our technique for fitting a blackbody to these four
PISN models before considering their visibility with the JWST. We then
estimate the total observable number in each case and for each
feedback prescription.  Finally, we briefly discuss the challenges
involved in actually identifying PISNe as such.

\subsection{A Simple Lightcurve Model}
\begin{figure}
  \begin{center}
    \includegraphics[width=\columnwidth]{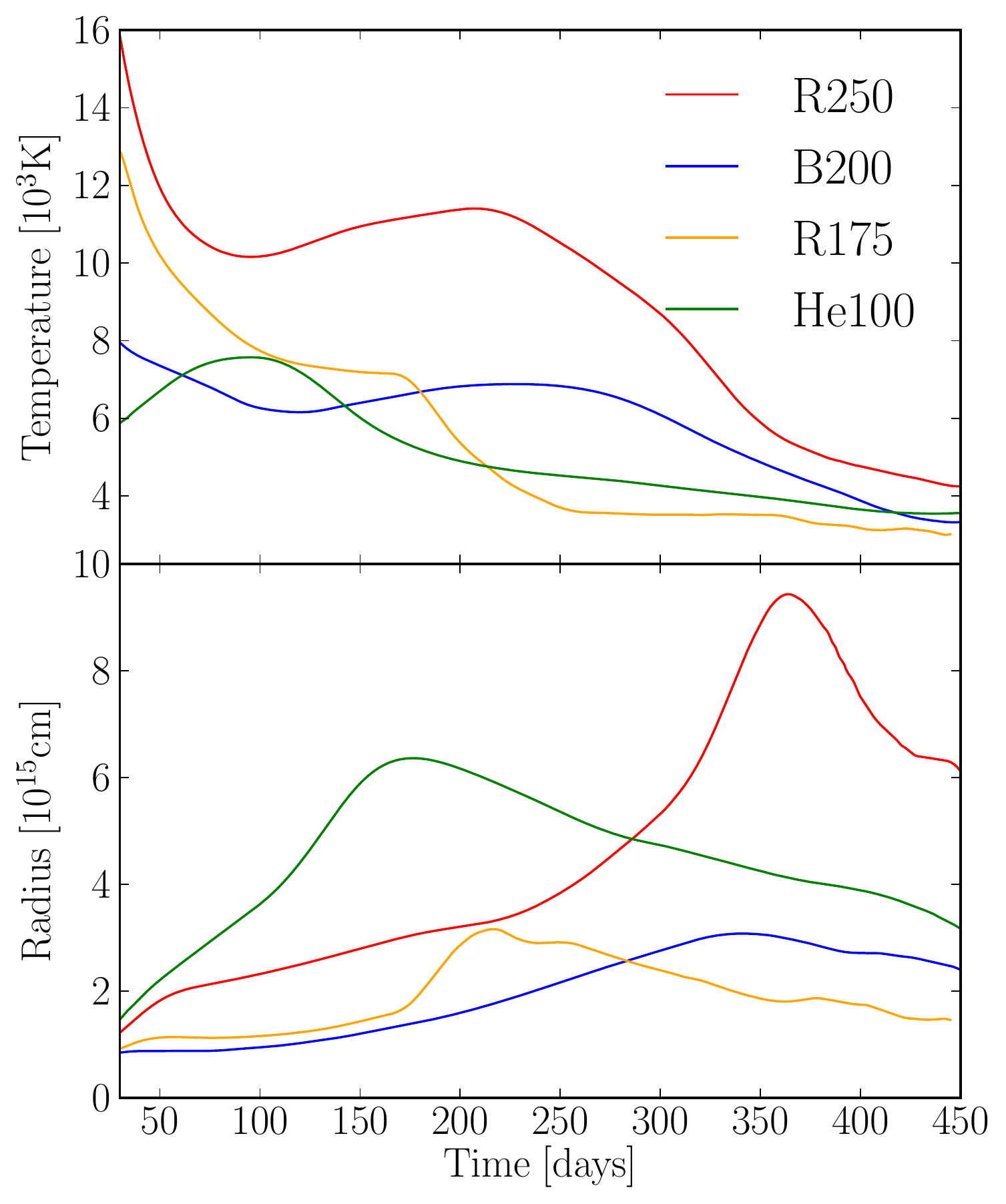}
    \caption{\footnotesize The temperature and radius of our blackbody
      fits as a function of source frame time since the explosion
      for models R250, B200, R175 and He100.  The secondary rise in
      temperature seen in R250 and B200 is caused by the decay of
      $^{56}$Ni reheating the ejecta at late times.  Note also that the
      apparent radii begin to decrease again at late times; this can
      be interpreted as the photosphere receding into the ejecta as
      the material cools.}
    \label{bbmodels}
  \end{center}
\end{figure}

In order to determine how long a PISN will be visible we must model
the source spectrum. Given the large mass involved, the ejecta will
remain optically thick until late times, so we make the reasonable
assumption that the PISN emits as a blackbody for the majority of its
visible lifetime.  Using the U, B, V, R, I, J, H, and K absolute
magnitude light curves presented in \citet{KasenWoosleyHeger2011}, we
perform a least-squares fit to find the combination of temperature $T$
and radius $R_{\textsc{sn}}$ that best matches the broadband
magnitudes at each point in time. This is done with the assumption
that the specific luminosity of the PISN $L_{\textsc{sn}, \lambda}$ at
wavelength $\lambda$ is given by
\begin{equation}
L_{\textsc{sn},\lambda} = 4\pi^2 R_{\textsc{sn}}^2 B_{\lambda}(T),
%= 4\pi R_{\textsc{sn}}^2 F_{\lambda} 
\end{equation}
where $B_{\lambda}$ is the Planck function. The resulting fits for the
evolution of the temperature and radius of the PISN models are shown in
\RefFig{bbmodels}.
Note that our blackbody assumption breaks down at late times when the
photosphere begins to recede into the ejecta.  This is manifested as
an apparent decrease in the radius of the PISN remnant.

\begin{figure*}[t!]
 \begin{center}
   \resizebox{15cm}{12cm}
   {\unitlength1cm
     \begin{picture}(15,12)
       \put(0,6){\includegraphics[width=7.5cm,height=6cm]{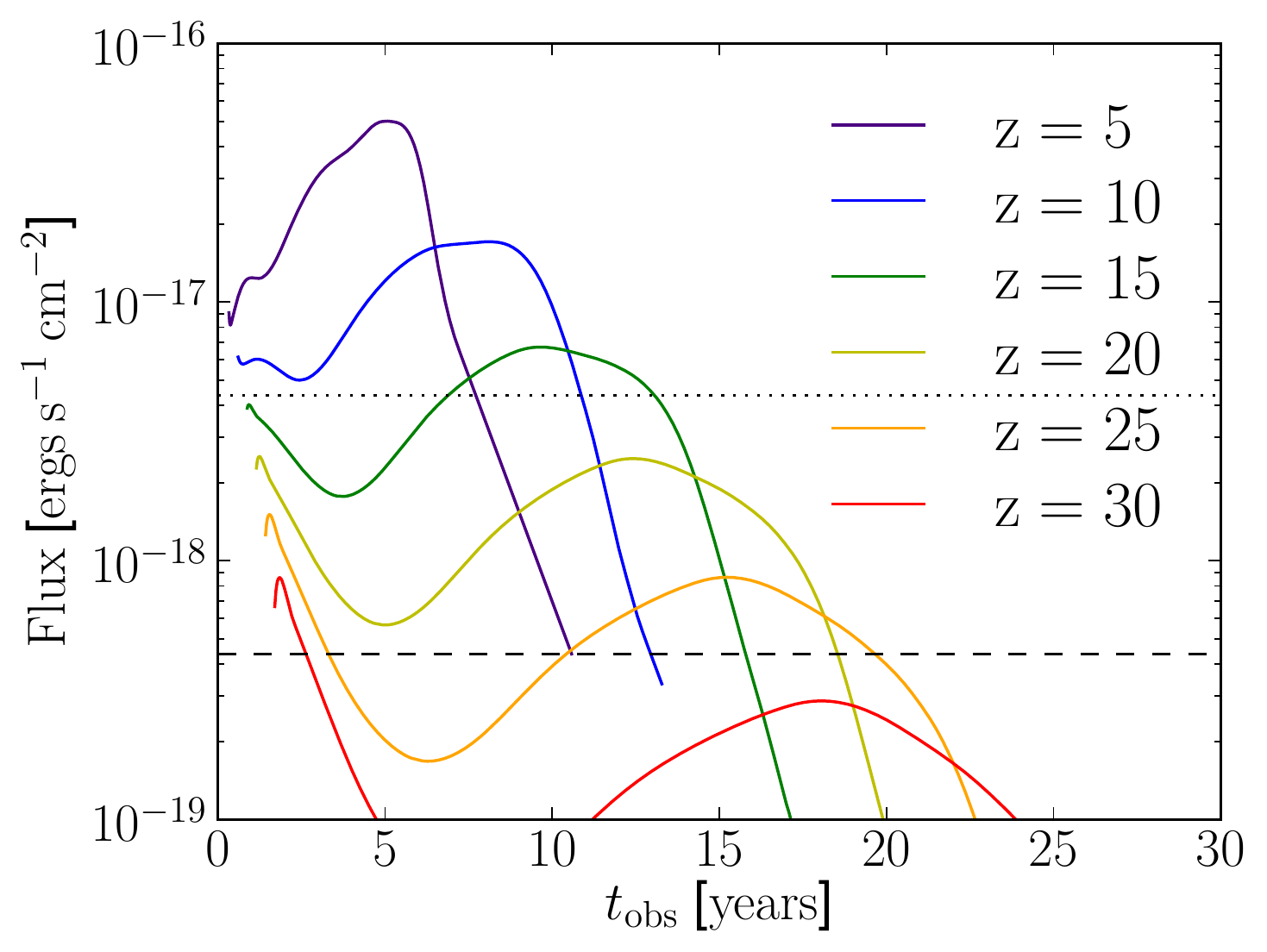}}
       \put(7.5,6){\includegraphics[width=7.5cm,height=6cm]{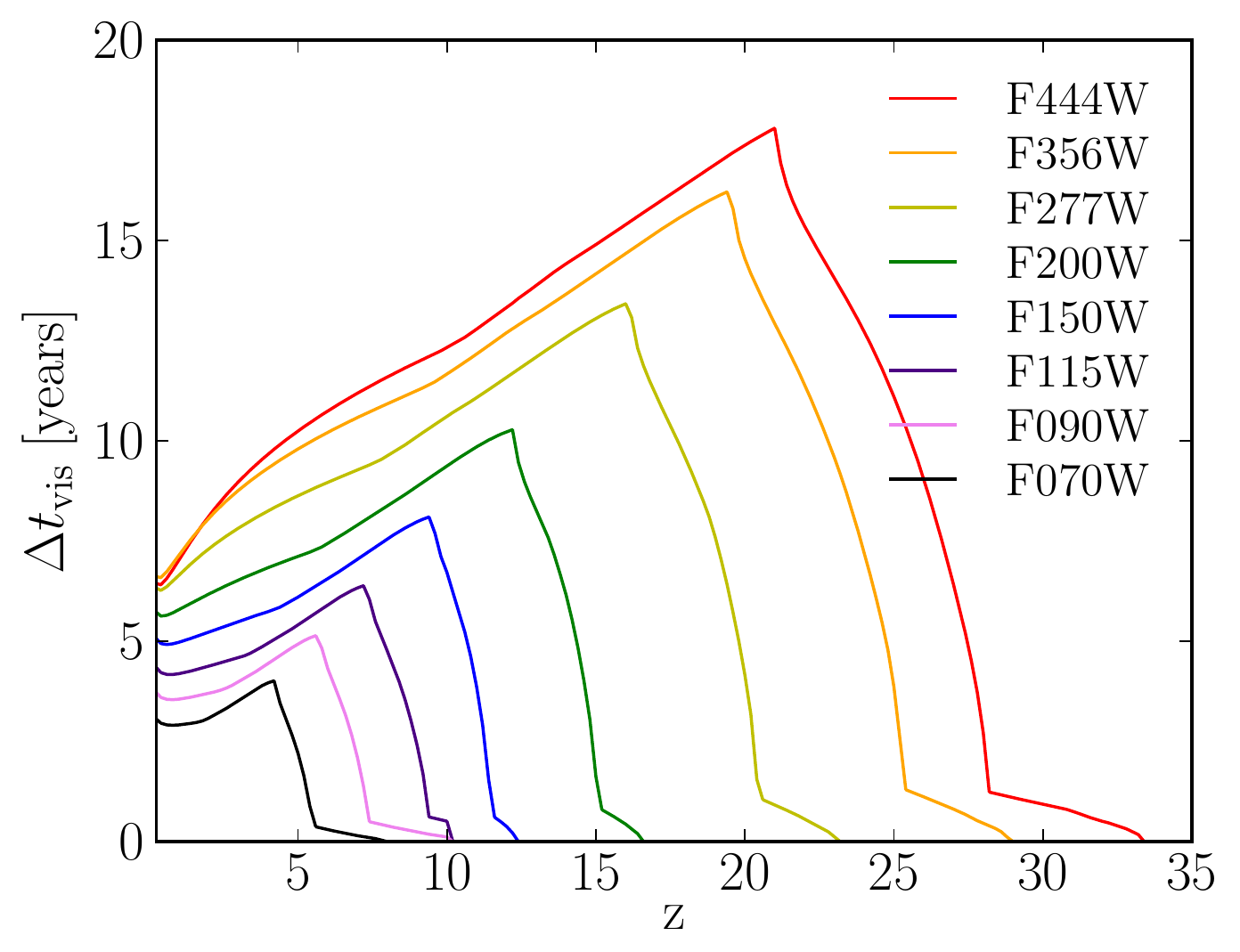}}
       \put(0,0){\includegraphics[width=7.5cm,height=6cm]{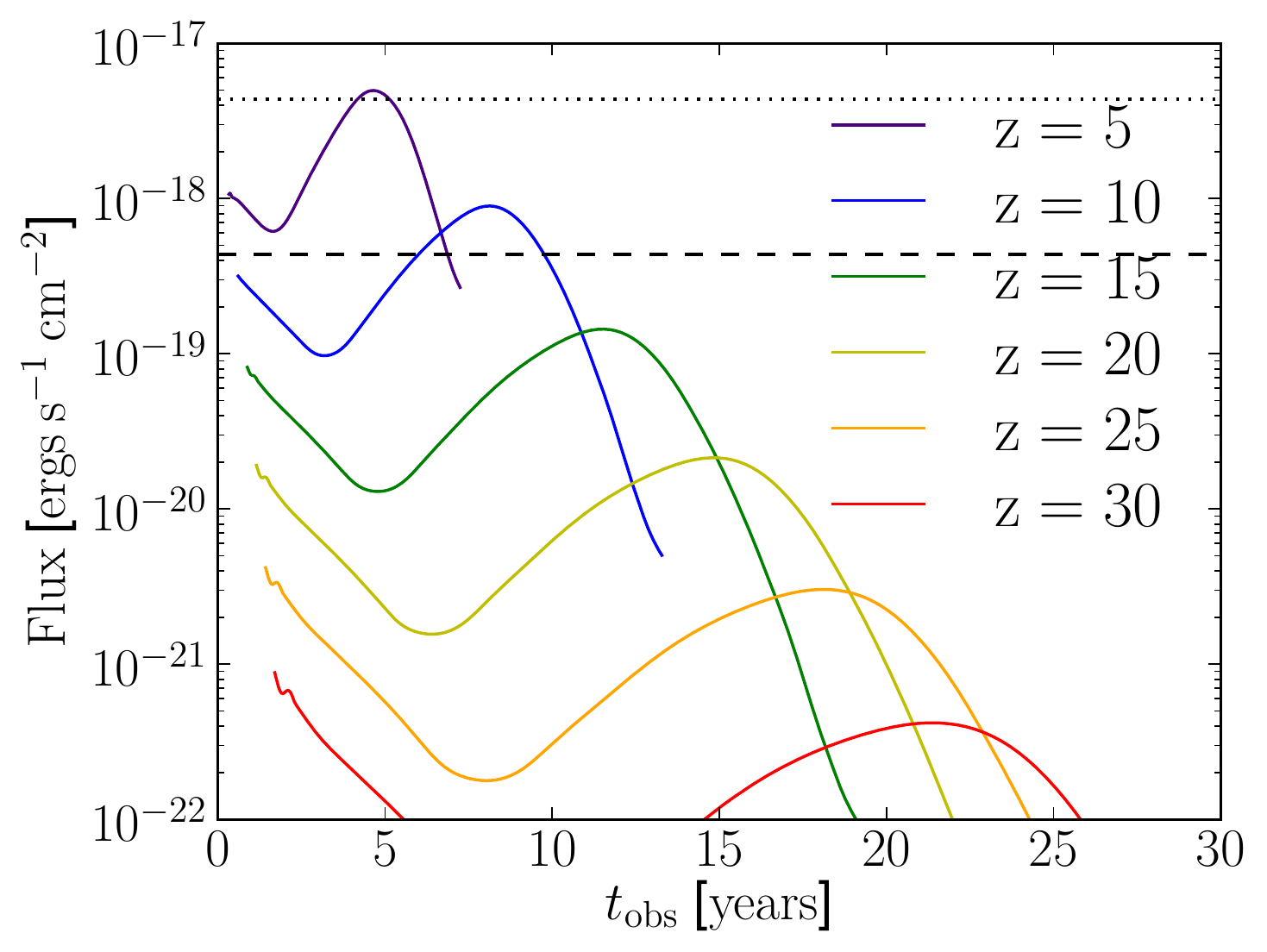}}
       \put(7.5,0){\includegraphics[width=7.5cm,height=6cm]{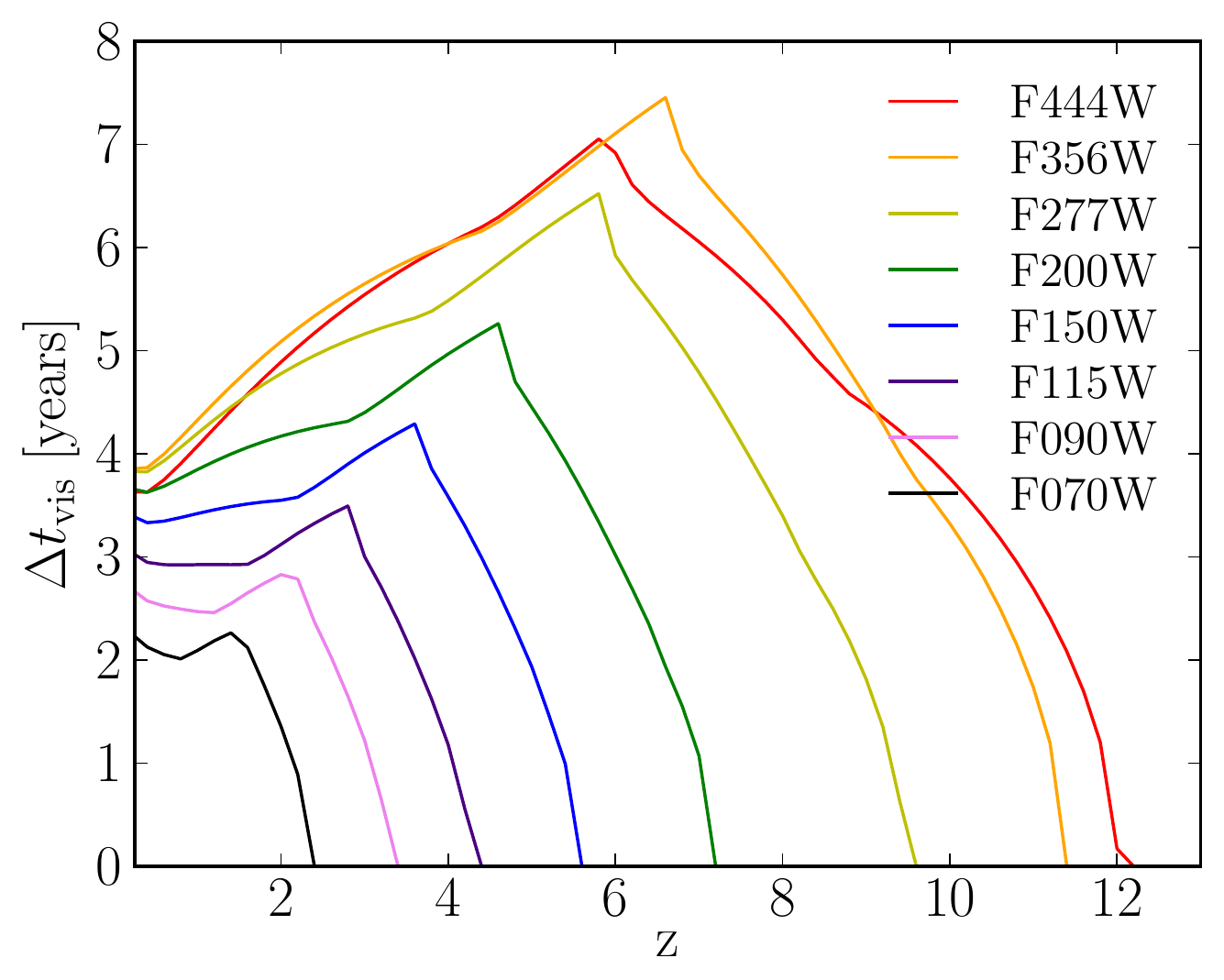}}
  \end{picture}}
\caption{\footnotesize Left: Lightcurves for the
  \citet{KasenWoosleyHeger2011} R250 (top) and B200 (bottom) models as
  they would be observed by JWST's F444W NIRCam filter at $z = 5, 10,
  15, 20, 25 \:{\rm and}\: 30$. The flux limits for a $10^6\,$s
  (dashed line) and $10^4\,$s (dotted line) exposure are shown for
  reference.  Right: The visibility time $\Delta t_{\rm vis}$ in years
  for R250 (top) and B200 (bottom) as a function of redshift for each
  of the NIRcam wide filters. Note that the axes are scaled
  independently.  Similar plots for models He100 and R175 are included
  in the appendix.}
   \label{visibility}
 \end{center}
\end{figure*} 

With this information we can then calculate the specific flux
$F_{\lambda, \rm em}$ in the rest frame and|accounting for redshift
and cosmological dimming|in the observer's frame for a source at
redshift $z$:
\begin{equation}
F_{\lambda, \rm obs}(T,R_{\textsc{sn}}, z) = \pi
\left[\frac{R_{\textsc{sn}}}{D_L(z)}\right]^2
\frac{B_{\lambda'}(T)}{1+z}.
\end{equation}
Here $\lambda' = \lambda/(1+z)$ accounts for redshifting, and the
luminosity distance $D_L = (1+z) r(z)$ accounts for cosmological
dimming.  Convolving this spectrum with a filter function
$\phi_\textsc{x}(\lambda)$ yields the observable flux in filter~X:
\begin{equation}
F_{\rm obs, \textsc{x}} = \int_{0}^{\infty} \phi_\textsc{x}(\lambda)
F_{\lambda, \rm obs}(T,R_{\textsc{sn}}, z) d\lambda.
\end{equation}

\subsection{Visibility}
The NIRCam instrument on the JWST will observe the early universe
through a number of narrow, medium-width, and wide
filters\footnote{http://www.stsci.edu/jwst/instruments/nircam/instrument-design/filters}.
The widest, longest-wavelength filter, F444W, will observe from 3.3 to
5.6 $\mu$m with a sensitivity limit of 24.5 nJy required for a
$10\sigma$ detection in $10^4\,$seconds \citep{Gardneretal2006}. Shown
in the left-hand column of \RefFig{visibility} is the observable flux
as it would appear in the F444W NIRCam filter at various redshifts for
the most and least easily observable models, R250 and B200,
respectively. See \RefFig{observability} for why these two were
chosen; models He100 and R175 can be found in the appendix.  The flux
limits for the filter of $4.4 \times 10^{-19}$ erg s$^{-1}$ cm$^{-2}$
for a $10^6\,$s exposure and $4.4 \times 10^{-18}$ erg s$^{-1}$
cm$^{-2}$ for a $10^4\,$s exposure are also shown for reference.  We
see that the brightest explosions (R250) would be visible to beyond
$z\sim25$, but are never so bright as to be detectable with current
generation telescopes.  This is consistent with the non-detection by
\citet{Frostetal2009} in a search of the {\it Spitzer}/IRAC Dark Field
for possible Pop~III PISN candidates.

To account for absorption of flux by neutral hydrogen along the line
of sight we implement a simple model of instant reionization at
$z=10$.  For sources above this redshift, we assume no flux is
observed shortward of the rest frame Ly$\alpha$ line.  This is not
relevant for the F444W NIRcam filter as Ly$\alpha$ does not redshift
into the filter until $z\sim40$, when the lightcurve is already far
below even the $10^6\,$s sensitivity limit.  It does however have an
effect, albeit a small one, on the F115W and F090W filters.

At low redshifts the duration of the lightcurve presented in
\citet{KasenWoosleyHeger2011} is not quite long enough for the
observed flux to reach the sensitivity limit; we extend it to the
limit by extrapolating assuming a power-law scaling.  The visible time
$\Delta t_{\rm vis}$ is then simply given by the time the lightcurve
is above the filter sensitivity limit.  Shown in \RefFig{visibility}
are the visibility times as a function of redshift for each of the
NIRcam filters.  

\begin{figure*}[ht!]
 \begin{center}
   \includegraphics[width=13cm]{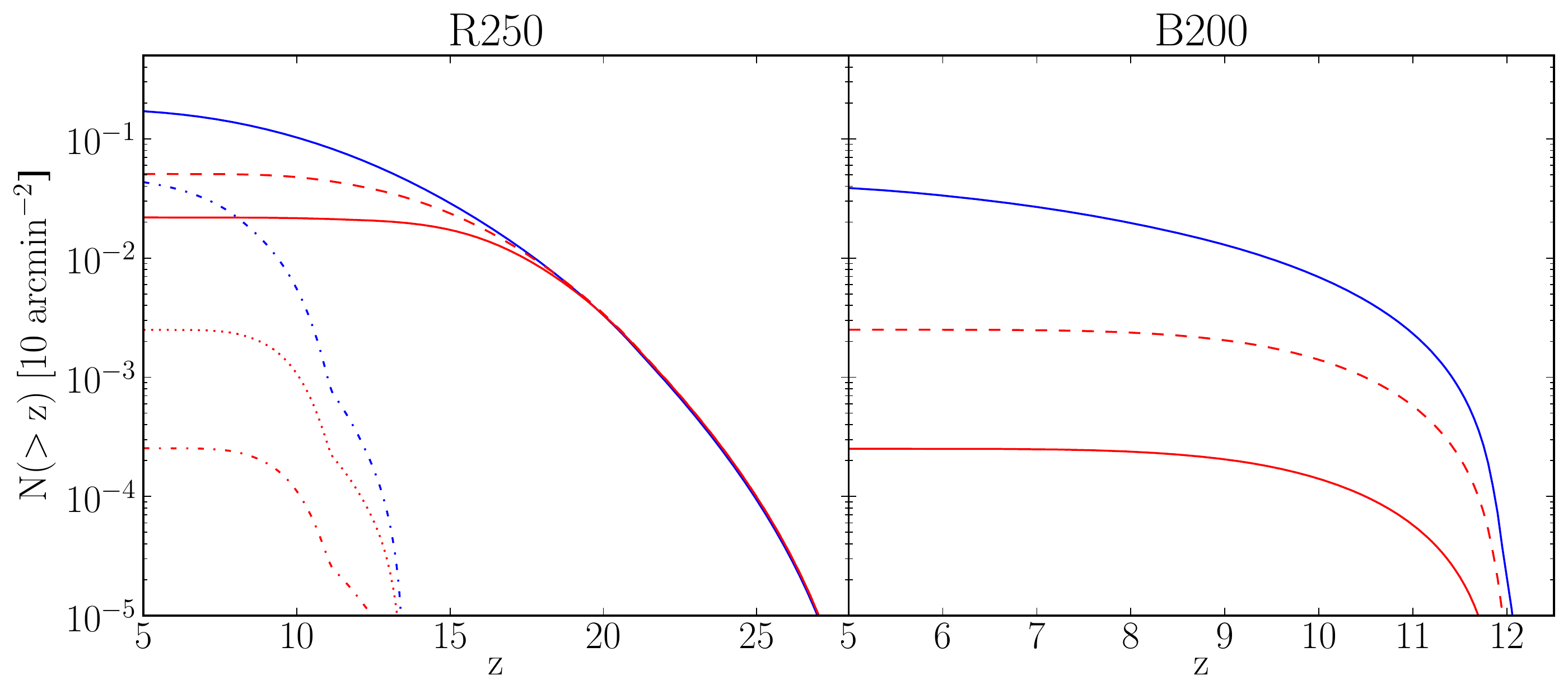}
   \caption{\footnotesize Upper and lower limits for the number of
     PISNe per JWST FoV above redshift $z$ with different feedback
     prescriptions. The observable numbers for a $10^6\,$s exposure
     assuming the R250 model are shown on the left; the B200 model is
     employed on the right. Solid blue lines show an upper limit to
     the observable number in the case of no feedback, solid red lines
     an estimate for the observable number in our conservative
     feedback scenario, and dashed red lines the number in the
     enhanced star formation case. Note that the x-axis is scaled
     independently in each panel.  Also shown in the R250 panel are
     the observable numbers for a $10^4\,$s exposure for the
     no-feedback (dot-dashed blue), conservative feedback (dot-dashed
     red), and enhanced star formation (dotted red) scenarios. The
     B200 model is not visible above $z=5$ in a $10^4\,$s exposure.}
   \label{obsnumber}
 \end{center}
\end{figure*} 
\subsection{The Observable Number}
 With this estimate for $\Delta t_{\rm vis}$, we may finally calculate
 the observable number of PISNe on the sky, given by the product of
 the PISN rate at $z$, as seen in the observer frame, and the time a
 PISN at $z$ is visible, $\Delta t_{\rm vis}$. This yields an estimate
 for the number of PISNe visible on the sky at any given time per unit
 redshift per unit solid angle:
\begin{equation} 
\frac{dN}{dz\,d\Omega} \simeq \frac{dN}{dt_{\rm
    obs}\,dz\,d\Omega}\,\Delta t_{\rm vis}.
\end{equation}
\RefFig{obsnumber} shows the number of PISNe per JWST field of view
(FoV) above redshift $z$ in a $10^6\,$s exposure for all three
feedback cases for models R250 and B200. Models He100 and R175 can be
found in the appendix.  For the R250 model the results for a $10^4\,$s
exposure are also included; B200 is not visible at high redshifts in a
$10^4\,$s exposure.

In the optimistic case of an R250-type PISN with no feedback we expect
\about0.2 PISNe per JWST FoV for a $10^6\,$s exposure.  In the most
pessimistic case of a B200-type PISN with strong negative feedback
this number drops to \about2.5$\,\times10^{-4}$ per FoV.  The actual
number detected by the JWST will most likely lie somewhere within this
range. Given this, we conclude that a single deep pencil-beam survey
is unadvisable for detecting PISNe, as there aren't enough in a given
field to ensure a detection, even in the most optimistic upper limit.
This suggests that a mosaic search, covering a larger area with shorter
exposure times, may be the best approach to ensure finding a Pop~III
PISN.

\subsection{PISN Identification}
The exceedingly long duration of their lightcurves poses a serious
challenge for identifying PISNe.  When combined with the cosmological
time dilation factors involved, PISN lightcurves can last for decades;
the highest redshift events will last longer than the projected
mission lifetime for the JWST, making the detection of PISNe by
searching for transients difficult at best.  However, a multi-year
campaign might be able to detect photometric variations; for example,
the He100 model would appear to decline in brightness by \about0.3
magnitudes per year at redshift $z=10$
\citep{KasenWoosleyHeger2011}. Additionally, PISN colors become redder
over time as the photosphere receeds into the ejecta and metal line
blanketing suppresses flux in the bluer bands.  While likely
insufficient to unambiguously identify PISNe, this could be useful in
selecting candidates for spectroscopic follow-up.

While the peak bolometric luminosities and spectra of PISNe resemble
those of typical Type Ia and Type II supernovae
\citep{JoggerstWhalen2011, KasenWoosleyHeger2011}, relatively little
mixing occurs during the explosion \citep{JoggerstWhalen2011,
  ChenHegerAlmgren2011}.  As a result, PISNe mostly retain their
onion-layer structure during the explosion, and metal lines do not
appear in the spectrum until late times when the photosphere has
receeded deep into the ejecta.  Some lighter elements may appear, but
the early spectrum of a PISN will be devoid of Si, Ni and Fe lines
\citep{JoggerstWhalen2011}. This may provide the spectroscopic
signature needed to identify PISNe.

\section{Discussion and Conclusions}
\label{conclusions} 
\begin{figure*}[t!]
\begin{center}
  \includegraphics[width=15cm]{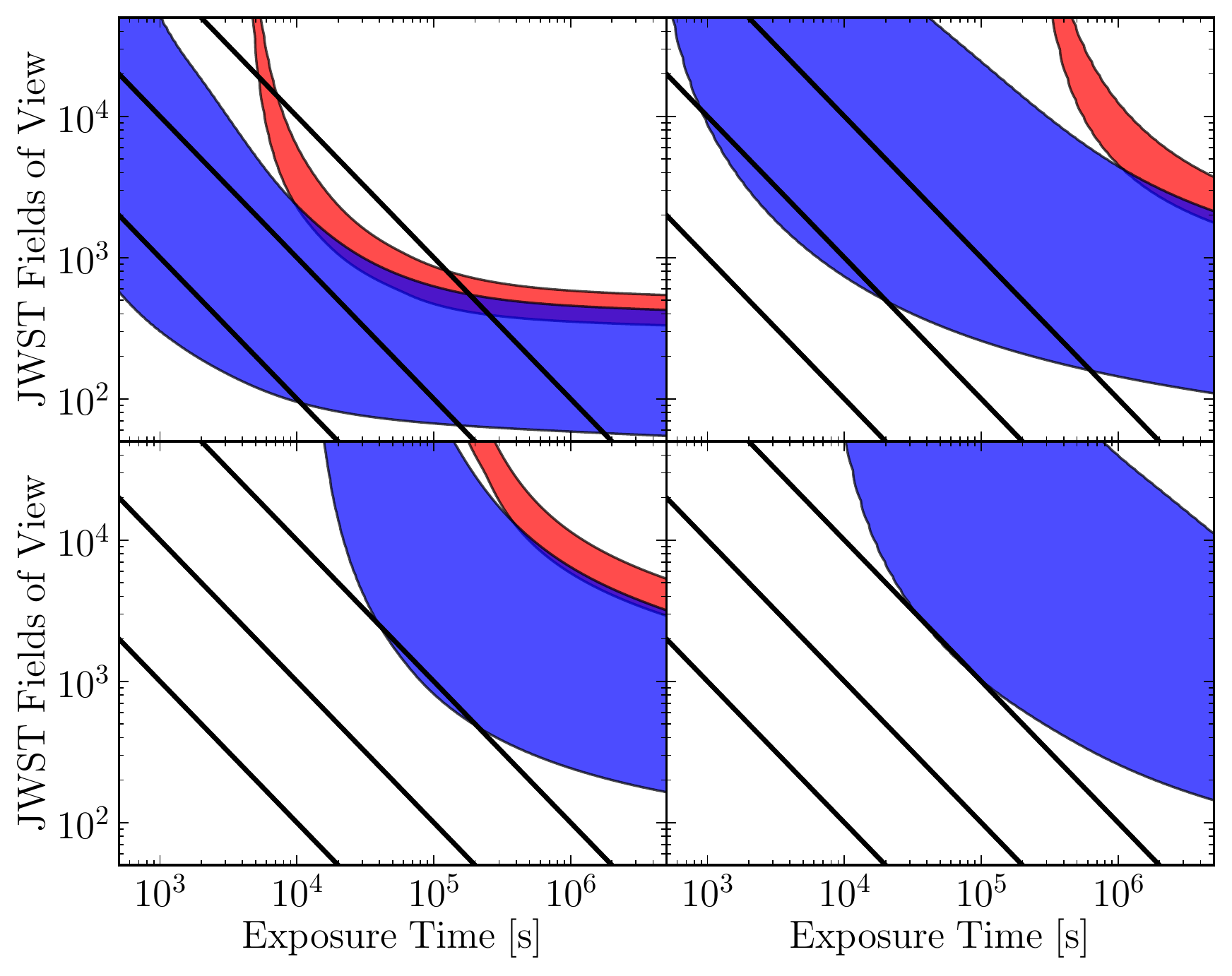}
  \caption{\footnotesize The observability of the R250, He100, B200,
    and R175 models (clockwise from upper left) from
    \citet{KasenWoosleyHeger2011} using the JWST's NIRcam F444W
    filter. Shown is the possible range for the number of JWST FoVs
    required to detect 10 sources as a function of exposure time. The
    blue range is for all PISNe, and the red for PISNe
    from $z>15$.  The lower boundaries correspond to the
    no-feedback upper limit to the PISN rate and the upper boundaries to
    the conservative feedback rate.  From left to right, the black
    lines represent the number of pointings possible in a total of
    $10^6$, $10^7$, and $10^8\,$s for a given exposure time.}
  \label{observability}
\end{center}
\end{figure*}

In this work, we have examined the source density of PISNe from
Pop~III stars and considered their detectability with the JWST. We
conclude that the limiting factor in detecting PISNe will be the
scarcity of sources rather than their faintness, in agreement with the
conclusions of \citet{WeinmannLilly2005}. The brightest PISNe should
be readily detectable with the longest wavelength NIRcam filters out
to $z\sim25$; the problem is the overall scarcity of sources.

 We have derived an estimate for the observable PISN rate, finding an
 upper limit of just over 0.01 PISNe per year per JWST FoV in the case
 of negligible chemical and radiative feedback. We also find that the
 inclusion of feedback can reduce the PISN rate by an order of
 magnitude, to \about0.001 per FoV. Accounting for the possibility of
 enhanced massive star formation in halos affected by LW radiation improves
 this rate slightly, to \about0.003 per FoV.  The derived PISN rate
 then allows us to place an upper limit on the observable number of
 PISNe in a $10^6\,$s exposure of \about0.2 PISNe per JWST FoV in the
 no-feedback case. The most pessimistic case of a B200-type PISN with
 strong negative feedback reduces this number to
 \about2.5$\,\times10^{-4}$ per FoV, or one PISN per 4000 JWST fields
 of view.

 The long duration of PISN lightcurves imply that spectroscopic
 follow-up of PISNe will likely be of great importance. PISN
 lightcurves can last for decades when combined with the cosmological
 time dilation factors at high redshifts, making the detection of
 PISNe by looking for transients untenable.  However, a multi-year
 campaign could identify candidates photometrically, and the lack of
 metal lines in the spectrum at early times could provide a
 spectroscopic signature for identification.

We find that the main obstacle to observing PISNe is the paucity of
sources.  Beyond a moderate exposure time of a few times $10^4\,$s,
the observability of bright PISNe is not a strong function of exposure
time and is instead controlled by the source density; this is evident
from \RefFig{observability}, where we have shown the number of JWST
FoVs required to detect 10 PISNe (blue) as a function of exposure time
for each of our PISN models.  The upper boundaries correspond to our
conservative estimate for the PISN rate in the presence of feedback,
and the lower boundaries to the estimated rate without feedback. We can
see that the decrease in the required number of JWST pointings slows
considerably beyond \about10$^5\,$s, hence a deep pencil-beam survey
would not be advisable in searching for PISNe. Even for only
high-redshift sources ($z>15$; red) the dependence on exposure time is
still minimal, being controlled by the lack of sources once the
required imaging depth is reached.

\begin{figure*}
\begin{center}
  \includegraphics[width=\textwidth]{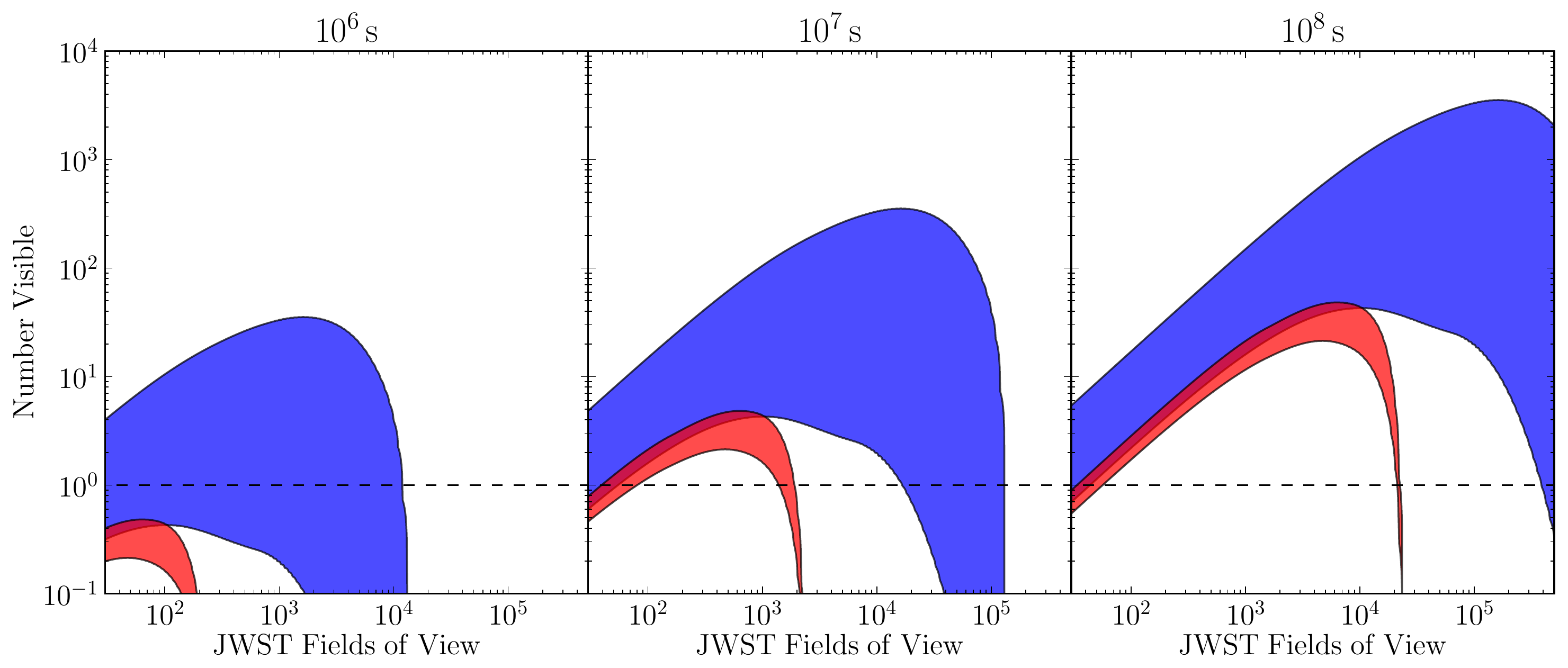}
  \caption{\footnotesize The total number of PISNe observable with a
    campaign of $10^6$, $10^7$ and $10^8\,$s (from left to right) as a
    function of survey area for the R250 PISN model.  In each case,
    the total campaign time is apportioned equally over the total
    survey area to determine the exposure time for individual
    pointings.  The blue region represents all PISNe, the red only
    PISNe from $z>15$.  Upper boundaries correspond to the no-feedback
    upper limit to the PISN rate and lower boundaries to the
    conservative feedback case. For reference we mark the case of only
    one PISN visible (dashed line).}
 \label{area_obsR250}
\end{center}
\end{figure*}  
Of particular interest in \RefFig{observability} are the black lines
representing the total number of pointings possible in $10^6$, $10^7$,
and $10^8\,$s for a given exposure time and their location relative to
the observability ranges in blue and red.  $10^6\,$s is approximately
the limit of what would be possible with a dedicated deep-field
campaign; $10^7\,$s is the limit of the observations the JWST could
make in a year assuming NIRCam is in use one third of the time;
$10^8\,$s (\about10 years) is the projected mission lifetime. 

While the detection of a PISN from a `first' star at very high
redshifts would be exciting and is in fact possible given the
detection limits of the JWST, the scarcity of sources at these
redshifts means that such a detection would be highly contingent on
serendipity. Even in the most optimistic case, with all available
minihalos producing an R250-type PISN, the observability range for
such events lies well above what is possible even in a full year of
observations, though a few may be detected over the lifetime of the
telescope.  The detection of a PISN at lower redshifts appears to be
more realistic.  As the faintest PISNe (R175 and B200) are effectively
unobservable, PISN searches should focus on looking for PISNe similar
to the R250 and He100 models.  In this case, the strategy with the
highest likelihood of detection will be a mosaic survey of many
moderately deep exposures.  This is clear from \RefFig{area_obsR250},
where we show the number of PISNe that will be observable with the
JWST in observing campaigns totalling $10^6$, $10^7$ and $10^8\,$s for
the R250 PISN model. The exposure time for each pointing varies with
the total area covered by the survey in order to keep the total
observing time constant.  Upper boundaries correspond to the number
visible in the no-feedback case, lower boundaries to the conservative
feedback case. As in \RefFig{observability}, the blue region shows the
observable number from all redshifts, the red region only those from
$z>15$.  We see that the observable number increases until the
resulting exposure time is no longer sufficient to detect PISNe.  The
optimal search strategy then will be to cover as large an area as
possible, going only as deep as necessary, possibly in a similar
manner to the ongoing Brightest of Reionizing Galaxies survey with the
{\it Hubble Space Telescope} \citep{Trentietal2011, Bradleyetal2012}.

\phantom{.} 

\acknowledgments{V.B.\ and M.M.\ acknowledge support from NSF grants
  AST-0708795 and AST-1009928 and NASA ATFP grant NNX09AJ33G. V.B.\
  thanks the Max-Planck-Institut f\"{u}r Astrophysik for its
  hospitality during part of the work on this paper. The simulations
  were carried out at the Texas Advanced Computing Center (TACC). }

\phantom{.}

\bibliography{references}

\begin{figure}[h!]
 \begin{center}
   \includegraphics[width=12.85cm]{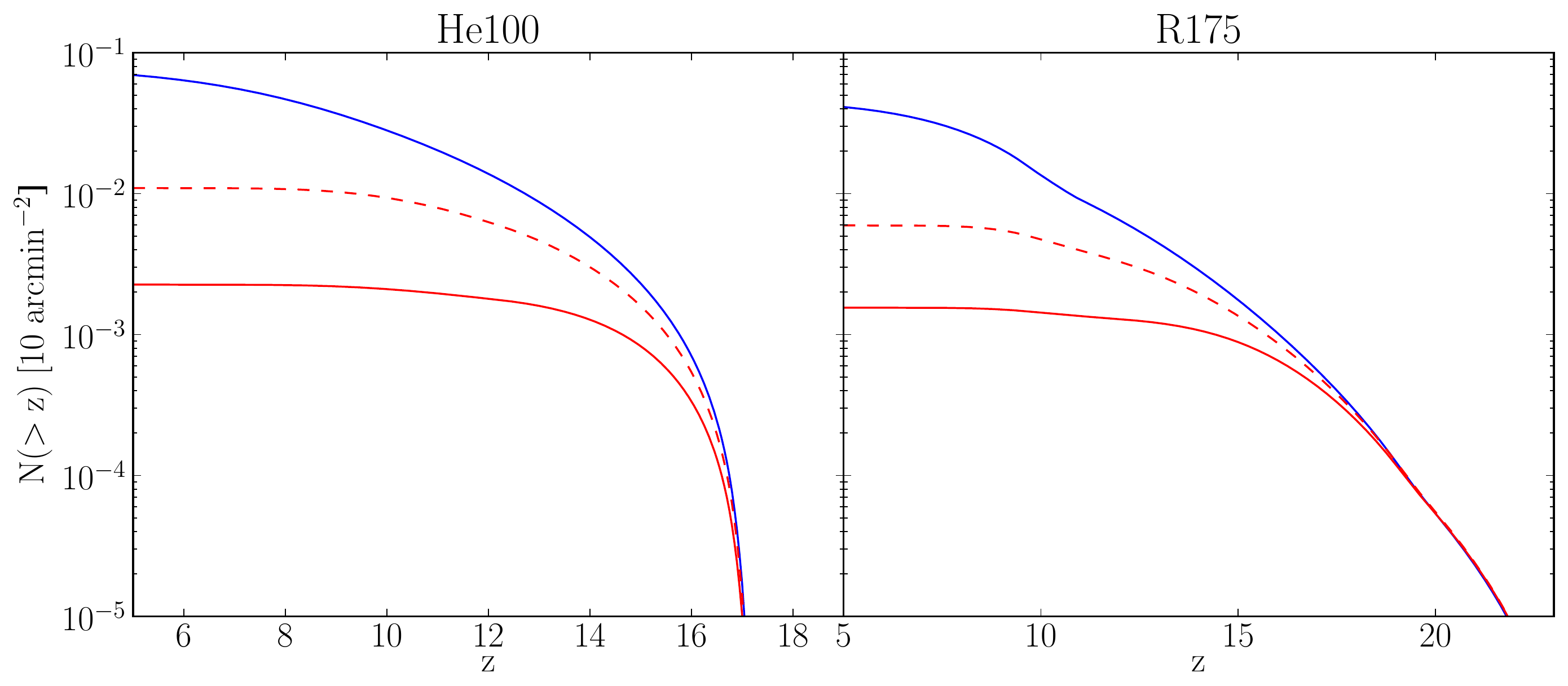}
   \caption{\footnotesize Upper and lower limits for the number of
     PISNe per JWST FoV above redshift $z$ with different feedback
     prescriptions. The observable numbers for a $10^6\,$s exposure
     assuming the He100 model are shown on the left; the R175 model is
     employed on the right. Solid blue lines show an upper limit
     to the observable number in the case of no feedback, solid red
     lines an estimate for the observable number in our conservative
     feedback scenario, and dashed red lines the number in the
     enhanced star formation case. Note that the x-axis is scaled
     independently in each panel.  }
   \label{obsnumber2}
 \end{center}
\end{figure}

\begin{figure}[p]
  \vspace*{\fill}
  \begin{center}
    \resizebox{15cm}{12cm}
              {\unitlength1cm
                \begin{picture}(15,12)
                  \put(0,6){\includegraphics[width=7.5cm,height=6cm]{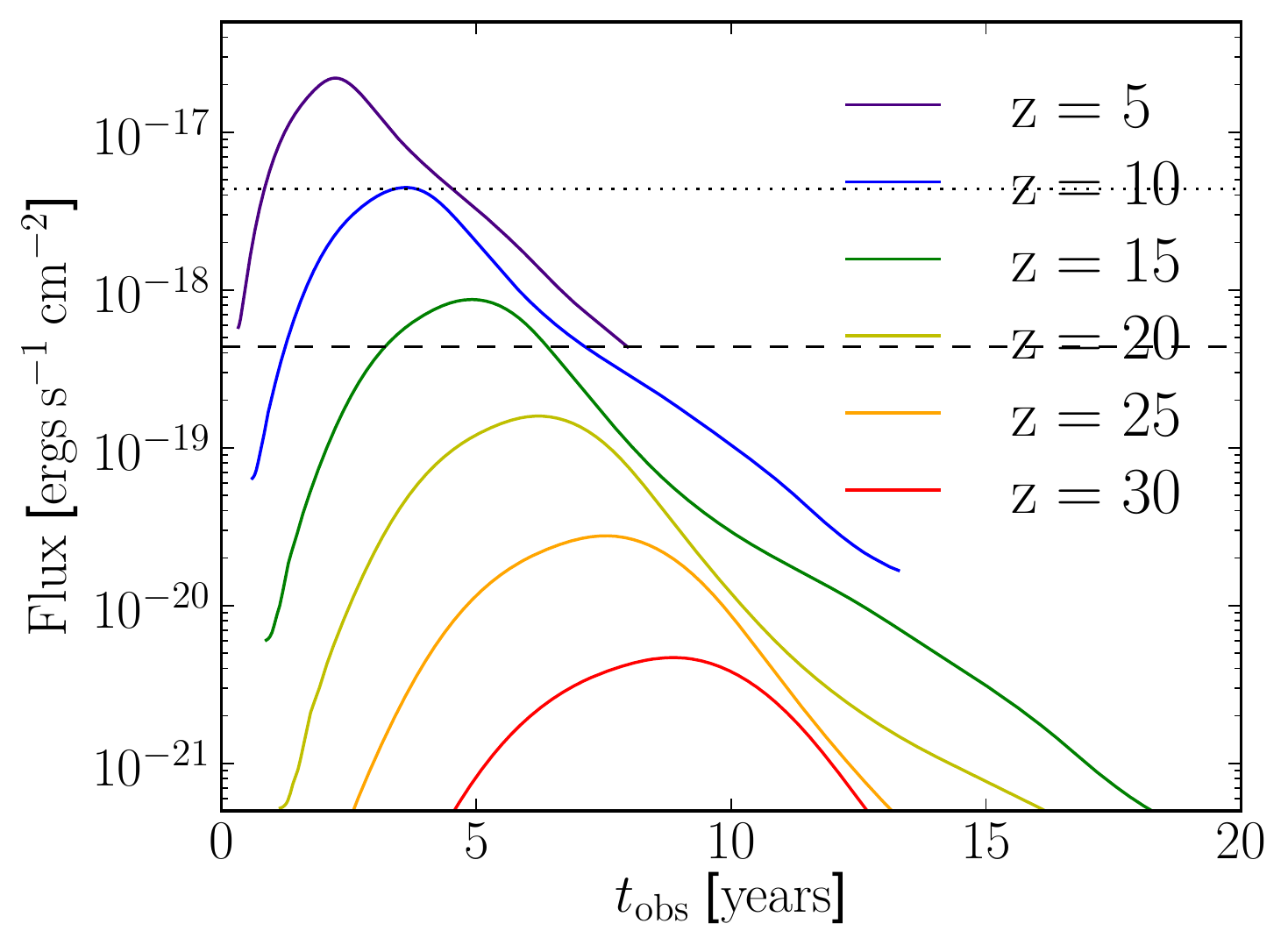}}
                  \put(7.5,6){\includegraphics[width=7.5cm,height=6cm]{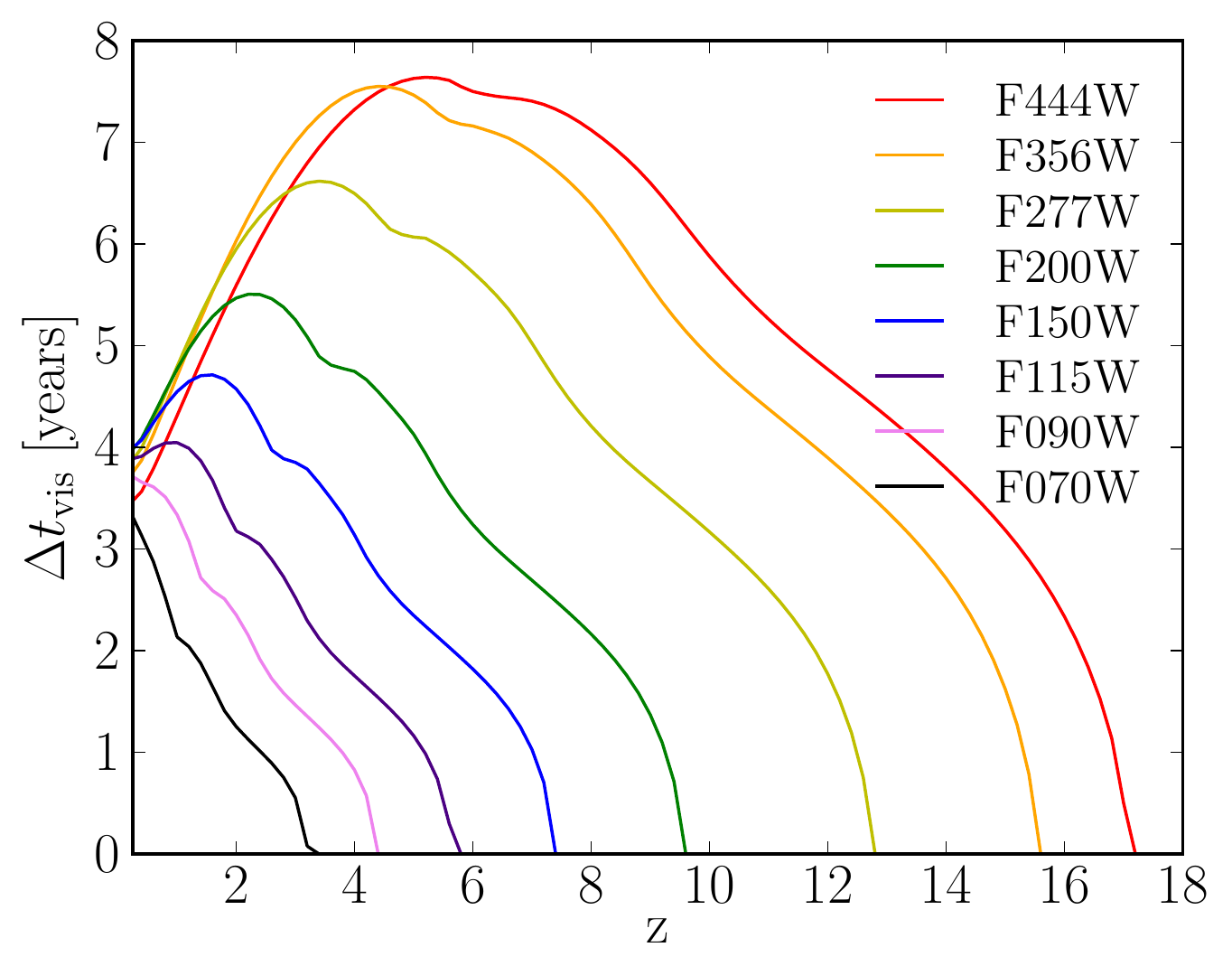}}
                  \put(0,0){\includegraphics[width=7.5cm,height=6cm]{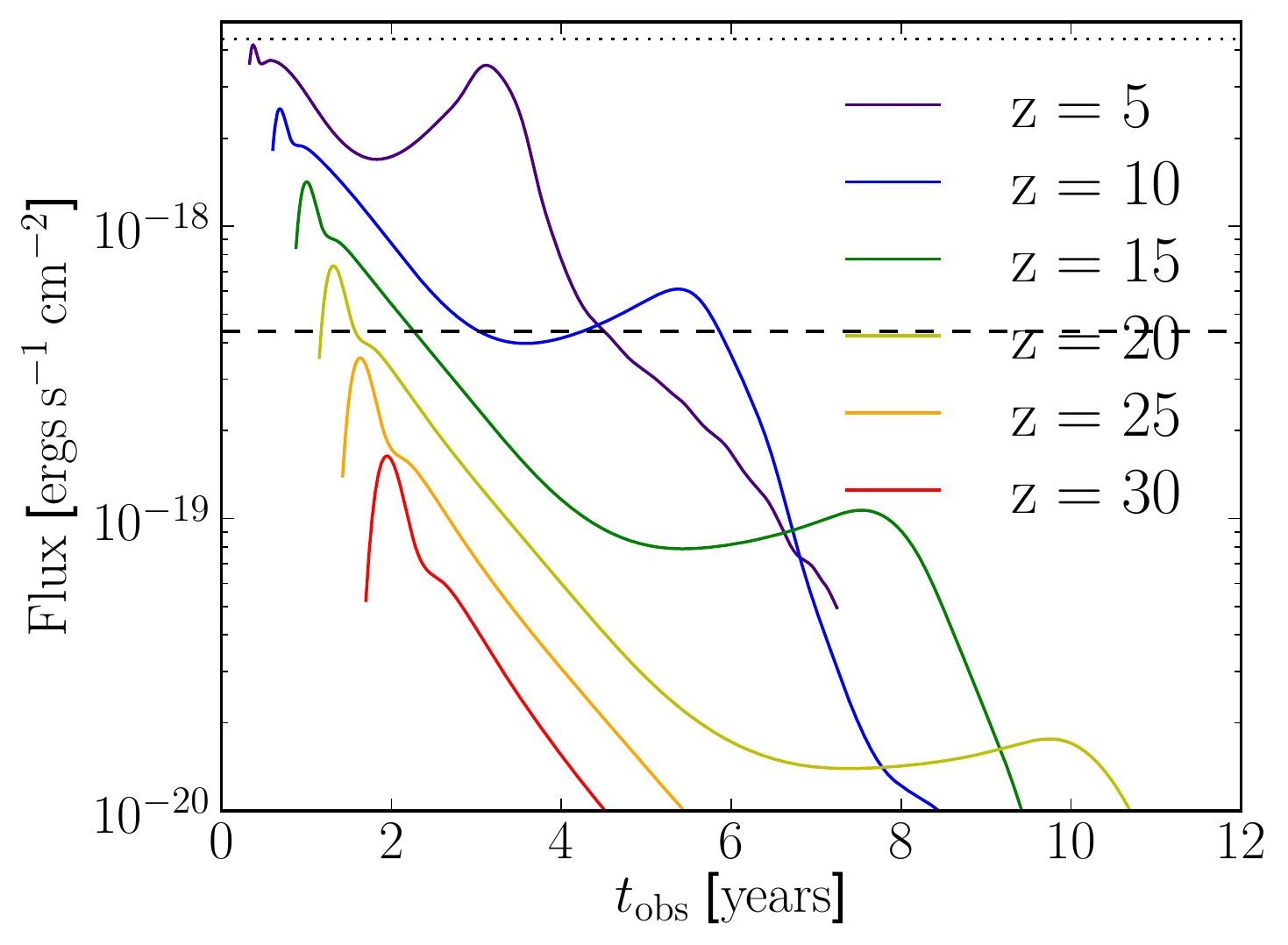}}
                  \put(7.5,0){\includegraphics[width=7.5cm,height=6cm]{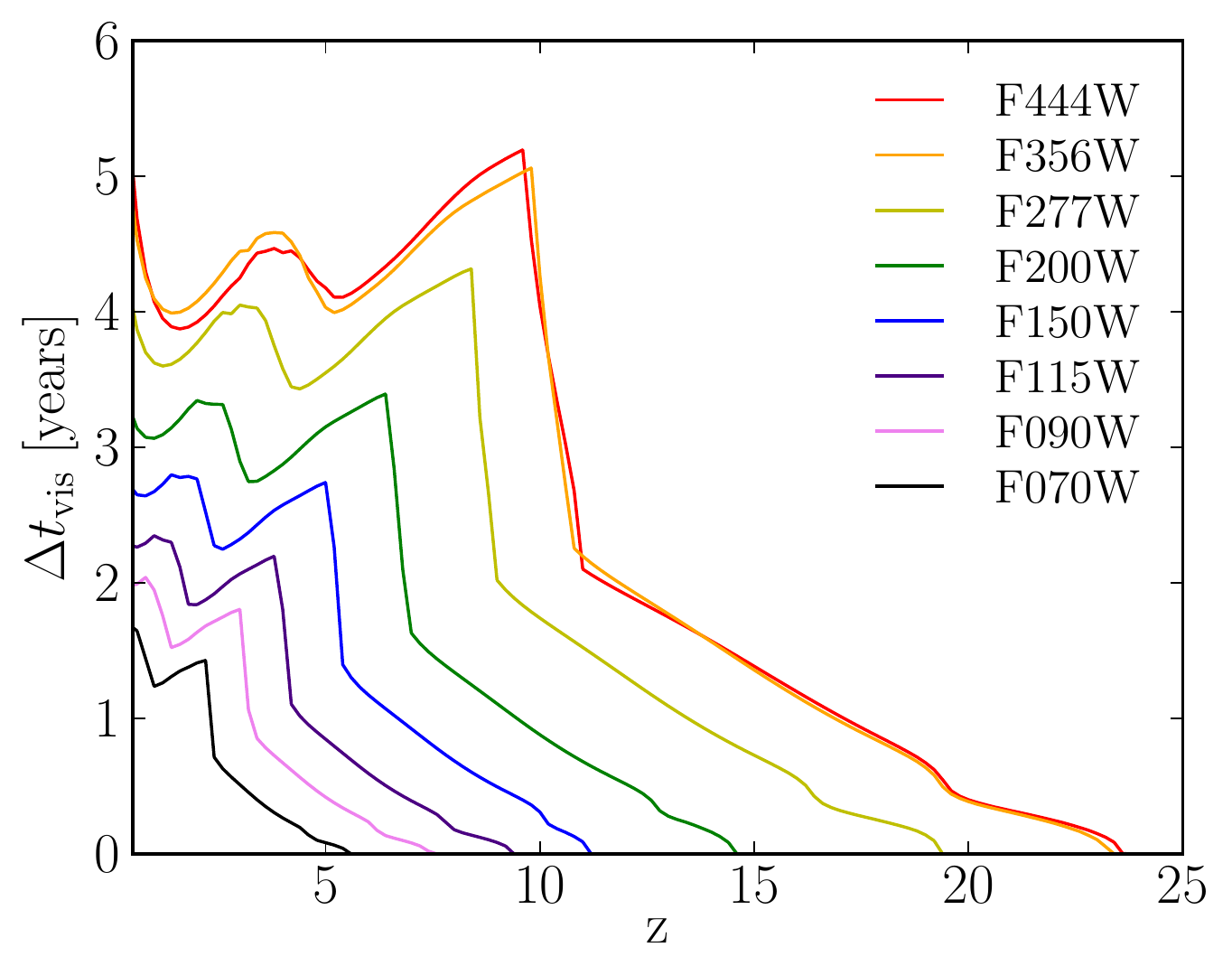}}
              \end{picture}}
              \caption{\footnotesize Left: Lightcurves for the
                \citet{KasenWoosleyHeger2011} He100 (top) and R175
                (bottom) models as they would be observed by JWST's
                F444W NIRCam filter at $z = 5, 10, 15, 20, 25 \:{\rm
                  and}\: 30$. The flux limits for a $10^6\,$s (dashed
                line) and $10^4\,$s (dotted line) exposure are shown
                for reference.  Right: The visibility time $\Delta
                t_{\rm vis}$ in years for He100 (top) and R175
                (bottom) as a function of redshift for each of the
                NIRcam wide filters. Note that the axes are scaled
                independently.}
              \label{visibility2}
  \end{center}
  \vspace*{\fill}
\end{figure}

\appendix
\label{appendix}
Included here are the observable properties of the He100 and R175 models from
\citet{KasenWoosleyHeger2011}, including the observable flux and time visible 
(\RefFig{visibility2}) and the observable number for each feedback scenario 
(\RefFig{obsnumber2}).  These plots are to be compared to
\RefFig{visibility} and \RefFig{obsnumber} in \RefSec{JWSTobs}.  Both
cases may be detected with NIRCam beyond $z=15$ for deep exposures of
$10^6\,$s.

\begin{figure}[p]
  \vspace*{\fill}
  \begin{center}
    \includegraphics[width=\textwidth]{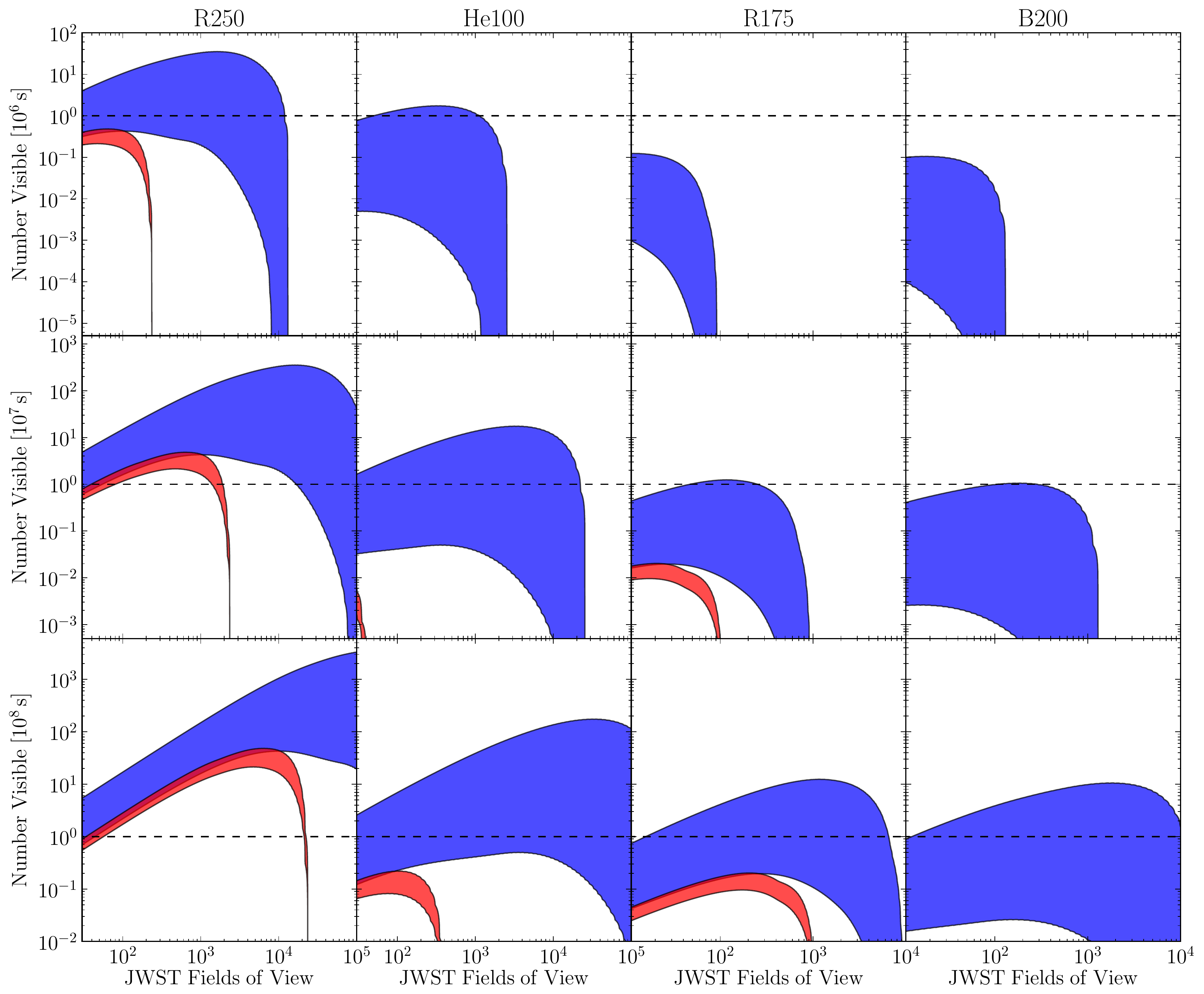}
    \caption{\footnotesize The total number of PISNe observable with a
      campaign of $10^6$, $10^7$ and $10^8\,$s (from top to bottom) as
      a function of total survey area for a given PISN model.  In each
      case, the total campaign time is apportioned equally over the
      survey area to determine the exposure time for individual
      pointings.  The blue region represents all PISNe, the red only
      PISNe from $z>15$.  Upper boundaries correspond to the
      no-feedback upper limit to the PISN rate and lower boundaries to
      the conservative feedback case.  The dashed line marks one PISN
      visible. Note that not all axes are scaled the same.  }
    \label{area_obs}
  \end{center}
  \vspace*{\fill}
\end{figure}

Also included is an extended version of \RefFig{area_obsR250} showing
the detectable number in observing campaigns totalling $10^6$, $10^7$
and $10^8\,$s for all four PISN models (\RefFig{area_obs}). Only
R250-type PISNe will be detectable above reshift 15, and R175- and
B200-type PISNe will only be detectable in observing campaigns
totalling greater than $10^7\,$s.

\end{document}